\documentclass[a4paper,11pt]{article}
\pdfoutput=1 % if your are submitting a pdflatex (i.e.\ if you have
\usepackage{jcappub} % for details on the use of the package, please
\usepackage{subcaption}
\usepackage[T1]{fontenc} % if needed
\usepackage{sidecap}
\usepackage{subcaption}
\usepackage{booktabs}
\usepackage{array}
\usepackage{cleveref}
\usepackage{xcolor}
\usepackage{soul}
\usepackage{amsmath, graphicx}
\usepackage{diagbox}

\title{\texttt{WarmSPy}: a numerical study of cosmological perturbations \\ in warm inflation}

\author[a,1]{Gabriele Montefalcone,\note{Corresponding author.}}
\author[a]{Vikas Aragam,}
\author[b,c]{Luca Visinelli}
\author[a,d,e]{and Katherine Freese}

\affiliation[a]{Texas Center for Cosmology and Astroparticle Physics, Weinberg Institute for Theoretical Physics, Department of Physics, University of Texas, Austin, Texas 78751, USA}
\affiliation[b]{Tsung-Dao Lee Institute (TDLI),
520 Shengrong Road, 201210 Shanghai, P.\ R.\ China}
\affiliation[c]{School of Physics and Astronomy, Shanghai Jiao Tong University,
800 Dongchuan Road, 200240 Shanghai, P.\ R.\ China}
\affiliation[d]{The Oskar Klein Centre, Department of Physics, Stockholm University, AlbaNova, SE-10691 Stockholm, Sweden}
\affiliation[e]{Nordic Institute for Theoretical Physics (NORDITA), 106 91 Stockholm, Sweden}

\emailAdd{montefalcone@utexas.edu}
\emailAdd{aragam@utexas.edu}
\emailAdd{luca.visinelli@sjtu.edu.cn}
\emailAdd{ktfreese@utexas.edu}

\abstract{We present \texttt{WarmSPy}, a numerical code in Python designed to solve for the perturbations' equations in warm inflation models and compute the corresponding scalar power spectrum at CMB horizon crossing. In models of warm inflation, a radiation bath of temperature $T$ during inflation induces a dissipation (friction) rate of strength $Q \propto T^c/\phi^m$ in the equation of motion for the inflaton field $\phi$. While for a temperature-independent dissipation rate ($c=0$) an analytic expression for the scalar power spectrum exists, in the case of a non-zero value for $c$ the set of equations can only be solved numerically. For $c>0$ ($c<0$), the coupling between the perturbations in the inflaton field and radiation induces a growing (decaying) mode in the scalar perturbations, generally parameterized by a multiplicative function $G(Q)$ which we refer to as the scalar dissipation function. Using \texttt{WarmSPy}, we provide an analytic fit for $G(Q)$ for the cases of $c=\{3,1,-1\}$, corresponding to three cases that have been realized in physical models. Compared to previous literature results, our fits are more robust and valid over a broader range of dissipation strengths $Q\in[10^{-7},10^{4}]$. Additionally, for the first time, we numerically assess the stability of the scalar dissipation function against various model parameters, inflationary histories as well as the effects of metric perturbations. As  a whole, the results do not depend appreciably on most of the parameters in the analysis, except for the dissipation index $c$, providing evidence for the universal behaviour of the scalar dissipation function $G(Q)$.}

\preprint{UTWI-25-2023, NORDITA-2023-031}

\begin{document}
\maketitle
\flushbottom

\section{Introduction}
\label{sec: introduction}

The most compelling solution to the horizon, flatness, and monopole problems is inflation~\cite{Guth:1980zm, Linde:1981mu, Albrecht:1982wi, Kazanas:1980tx, Starobinsky:1980te, Sato:1980yn, Mukhanov:1981xt, Linde:1983gd, Mukhanov:1990me}. By virtue of accelerated expansion, inflation ensures that when the process concludes, the Universe achieves a state of significant flatness, homogeneity, and isotropy at the largest observable scales.
In addition to resolving these fundamental issues, inflation offers a mechanism for generating the density fluctuations that ultimately give rise to the observed large-scale structures within the cosmic web. The distinctive pattern in the fluctuation spectrum is expected to align with the angular power spectrum observed in the cosmic microwave background (CMB) by various experiments, including recent data from the {\it Planck} satellite and BICEP/Keck Array~\cite{Planck:2018nkj, Planck:2018jri}, as well as measurements of small-scale CMB fluctuations obtained by the Atacama Cosmology Telescope~\cite{ACT:2020gnv, ACT:2020frw, ACT:2023dou} and the South Pole Telescope~\cite{SPT-3G:2021eoc, SPT-3G:2021wgf}. These fluctuations, originating from quantum-mechanical effects, are inherently adiabatic and are attributed to perturbations in a scalar field known as the {\it inflaton} field, which drives the inflationary expansion.

Conventional models of inflation involve a single scalar field slowly rolling down a nearly flat potential, inducing a quasi-de Sitter phase during which the Universe inflates superluminally.  Once the field nears the minimum of its potential, the vacuum energy converts to radiation, thereby reheating the Universe. 

A well-established alternative framework to conventional inflation is \textit{warm inflation} (WI), in which the inflaton is thermally coupled to a bath of radiation~\cite{Berera:1995ie, Berera:1996fm}, see refs.~\cite{Fang:1980wi, Moss:1985wn, Yokoyama:1987an} for earlier attempts. The inflaton continually sources the production of radiation.  The resulting radiation bath induces a dissipation (friction) term in the equation of motion for the inflaton field, and also alleviates the need for a separate reheating phase at the end of inflation. For reviews of WI see refs.~\cite{Berera:2008ar,Bastero-Gil:2009sdq, Kamali:2023lzq}.

The presence of a radiation bath and dissipation not only alters the background dynamics of the inflaton field but also its perturbations. Specifically, the addition of these thermal effects can have a significant impact on the primordial power spectrum.  Fluctuations produced in WI models are primarily thermal in origin, with quantum fluctuations being subdominant in the limit of a large dissipation rate between the inflaton and radiation sectors. Quantitative models of warm inflationary perturbations must therefore take into account the effect of this dissipation. The established approach in WI (which we also follow in this work) is based on the stochastic inflation formalism, in which short-wavelength dynamics backreact on the long-wavelength modes via quantum and thermal noise terms in their equations of motion~\cite{Ramos:2013nsa}. The perturbations' evolution is then determined by solving the fully coupled equations for all scalar, radiation energy density, and radiation momentum perturbations.  

Although an accurate calculation of the production of perturbations and the resulting primordial power spectrum in WI must be done numerically,
 one can gain insight by parameterizing the power spectrum analytically as~\cite{Hall:2003zp, Graham:2009bf, Bastero-Gil:2011rva, Ramos:2013nsa}:
\begin{align}
    \Delta_{\mathcal{R}}^{2}=\left(\frac{H^{2}}{2 \pi \dot{\phi}}\right)^{2}\left(1 + 2 n_{\mathrm{BE}}+\frac{2 \sqrt{3} \pi Q}{\sqrt{3+4 \pi Q}} \frac{T}{H}\right) G(Q),
\label{eq:defG}
\end{align}
where $\dot\phi$ is the inflaton's background velocity, $n_{\mathrm{BE}} = [\exp(H/T )-1]^{-1}$ is the Bose-Einstein distribution, $T$ is the radiation temperature,  $Q\equiv \Upsilon/(3H)$ is a dimensionless parameter defining the strength of the dissipation, $\Upsilon$ is the dissipation rate, and $H$ is the Hubble parameter during inflation (setting the scale of inflation); all quantities are evaluated at horizon crossing. The parenthetical terms correspond to the analytic power spectrum of a temperature-independent dissipation rate, and the multiplicative factor $G(Q)$ accounts for the growth or suppression of inflaton fluctuations due to its direct coupling with radiation for a temperature-dependent dissipation rate. Here, we refer to $G(Q)$ as the scalar dissipation function. Determining $G(Q)$ can only be done numerically by solving the full set of perturbation equations found in WI  and is computationally demanding, particularly for large values of $Q$.

In this work, we present—to our knowledge—the first open-source and publicly available code to compute $G(Q)$ and the full, numerical power spectrum for warm inflationary models.\footnote{The code is available open-source at \href{https://github.com/GabrieleMonte/WarmSPy.git}{github.com/GabrieleMonte/WarmSPy.git}} We present all analytic and numerical steps for this computation in a fully reproducible manner, with a goal of providing a clear guide for readers interested in performing their own warm inflationary computations. The novelty in our work is the public code as well as the extensive study done on the relationship between the scalar dissipation function $G(Q)$ in eq.~\eqref{eq:defG} and other model parameters. The  system of perturbation equations was previously well known; we specifically follow the approach of ~\cite{Ramos:2013nsa}.  
However, their solutions had not been studied numerically in such a comprehensive way as we do in this paper.  Further, our numerical analysis extends to higher values of $Q$ that had previously been studied.

Several codes that solve for the evolution of the inflaton field are already present in the literature, in the context of conventional (cold) inflation models without dissipation. Publicly available numerical packages that yield the power spectrum for  multi-field inflation models include {\tt Pyflation}~\cite{huston:2011vt}, \texttt{Inflation.jl}~\cite{rosati_robert_2020_4708348}, and {\tt MultiModeCode}~\cite{price:2014xpa}. Various codes have been implemented to numerically evaluate inflationary correlation functions for single-field models, including {\tt CppTransport}~\cite{seery:2016lko}, {\tt PyTransport}~\cite{mulryne:2016mzv}, see also refs.~\cite{dias:2016rjq, ronayne:2017qzn}, BINGO~\cite{hazra:2012yn}, the work by Chen {\it et al.}~\cite{chen:2006xjb}, and the work by Horner and Contaldi~\cite{horner:2013sea}. Our work differs from these previous installments because it is designed to solve the equation of motion for the inflaton field along with the radiation bath content predicted in WI. The code includes various models for the inflaton potential and the coupling of the inflaton field to the radiation bath. The output generates the fitting formulas for the scalar dissipation function $G(Q)$.

We present our work as follows. Section~\ref{sec:background_ev} summarizes the background dynamics of WI and presents the notation used. In section~\ref{sec:perturbations}, we introduce the equations for the perturbations including the stochastic inflation formalism; we provide a simple proof to obtain the analytic expression that approximates the power spectrum of WI (eq.~\eqref{eq:defG}), and define the scalar dissipation function $G(Q)$. The various modules that compose our code \texttt{WarmSPy} are described in detail in section~\ref{sec:code}. The results for the function $G(Q)$ using different models and inflationary histories are compared in section~\ref{sec:results}, where we also propose a new fitting function that extends currently available fits in the literature to wider range of values of $Q$, of relevance in some WI constructions in the literature~\cite{Das:2020xmh,Berghaus:2019whh,Bastero-Gil:2019gao}.

\section{Background evolution in warm inflation
\label{sec:background_ev}}

In WI, the inflaton continually converts into radiation during inflation itself, as opposed to only reheating the universe once inflation ends. This dissipation mechanism is parameterized by the introduction of a non-negligible dissipation rate $\Upsilon = \Upsilon(T,\phi)$ that couples the inflaton field to the radiation energy density $\rho_r$ in their equations of motion ~\cite{Berera:1995wh, Berera:1995ie}:
\begin{eqnarray}
    \Ddot{\phi}+(3H+\Upsilon)\dot{\phi}+V_{,\phi} &=& 0\,,\label{eq:WI.1}\\
    \Dot{\rho}_R+4H\rho_R &=& \Upsilon \Dot{\phi}^2\,,  \label{eq:WI.2}
\end{eqnarray}
where a dot denotes the derivative with respect to cosmic time and $V_{,\phi}=\partial V/\partial \phi$. Conservation of the total energy of the system imposes that the energy lost by the inflaton field must be gained by the radiation's energy density, with the term on the RHS of eq.~\eqref{eq:WI.2} encapsulating this process.

The dissipation rate $\Upsilon$ describes the rate at which the energy in the inflaton field converts into radiation, and the detailed form of this term is model-dependent~\cite{Berera:1998gx, Yokoyama:1998ju}. Here, we parametrize the dissipation rate as:
\begin{equation}
    \Upsilon(\phi, T)=C_\Upsilon M^{1-c+m} \frac{T^c}{\phi^m}\,,
    \label{eq:Upsilon}
\end{equation}
where $C_\Upsilon$ is a dimensionless constant that depends on the underlying microphysical model, $M$ is a mass scale specified by the model, while $c$ and $m$ are integers that can be either positive or negative. Successful models of WI have been built by coupling the inflaton with the radiation bath indirectly through heavy mediators~\cite{Berera:2002sp, Moss:2006gt, Berera:2008ar} and directly by protecting the flatness of the inflaton potential from thermal and quadratic corrections~\cite{Bastero-Gil:2016qru, Bastero-Gil:2019gao, Berghaus:2019whh}. A discussion on the derivation of the dissipation coefficient using field theory tools is given in \cite{Bastero-Gil:2010dgy}.

In the following, we will work with $c=\{3,1,0,-1\}$, motivated by the explicit constructions of dissipative rates in the literature. For example, a decay process of the inflaton into light fermions mediated by a heavy boson field predicts $\Upsilon \propto T^3$~\cite{Moss:2006gt}. Further, $c=3$ also results  when considering an axion field coupled to a non-Abelian gauge group, with the friction term related to the sphaleron transition rate obtained from lattice gauge theory~\cite{Moore:2010jd, Laine:2016hma}. In the presence of a small mass to the final fermionic states, chirality-violating scattering suppresses the friction from the sphaleron transitions leading to $\Upsilon \propto T$~\cite{Berghaus:2020ekh}. Moreover, in the warm little inflaton model, where the inflaton field is still taken to be a pseudo Nambu-Goldstone boson, Yukawa coupling to a few light fields have been shown to produce a dissipation coefficient
with linear temperature dependence, i.e.\ $\Upsilon\propto T$~\cite{Bastero-Gil:2016qru}, and in the high temperature regime an inverse
temperature dependence, i.e.\ $\Upsilon \propto T^{-1}$~\cite{Bastero-Gil:2019gao}. 

The strength of the dissipation is typically parametrized by the dimensionless ratio $Q$:
\begin{equation}
    Q\equiv \frac{\Upsilon}{3H}.
\end{equation}
For $Q\gg1$, a strongly dissipative regime is achieved, while $Q < 1$ represents the weak regime of WI. In all cases, WI requires $T>H$, which is roughly the criterion for which thermal fluctuations dominate over quantum fluctuations~\cite{Berera:1995ie}. Throughout this work, we also assume that the radiation thermalizes on a time scale much shorter than $1/\Upsilon$~\cite{Berera:1995wh,Berera:1995ie}, so that the energy density of radiation can be taken to be:
\begin{equation}
    \rho_R(T)=\alpha_1 T^4, \quad \text{with } \quad \alpha_1=\frac{\pi^2}{30}g_{*}(T), \label{eq:WI.3}
\end{equation}
where $g_{*}(T)$ is the number of relativistic degrees of freedom for a radiation bath at temperature $T$.\footnote{Although we do not specify $g_{*}(T)$ in the equations, we adopt the number of relativistic degrees of freedom in the minimal supersymmetric Standard Model $g_{*}(T) = 228.75$ when presenting our results.}

The Hubble expansion rate $H\equiv \dot{a}/a$ is determined by the Friedmann equation, which reads:
\begin{equation}
    H^2=\frac{1}{3M^2_{\rm{pl}}}\left( V(\phi)+\frac{\dot{\phi}^2}{2}+\rho_r\right), \label{eq:WI.4}
\end{equation}
where $M_{\mathrm{pl}}= 1/\sqrt{8 \pi G} \, \approx 2.436\times 10^{18}\,\mathrm{GeV}$ is the reduced Planck mass.
Inflation takes place when the potential $V$ is approximately flat and the potential energy
dominates, such that the Hubble expansion rate $H$ is approximately constant. During this period, known as the slow-roll regime of the inflaton field, higher order derivatives in eqs.~\eqref{eq:WI.1} and~\eqref{eq:WI.2} can be neglected, i.e.
\begin{equation}
    \ddot{\phi} \ll H \dot{\phi}, \quad \text { and } \quad \dot{\rho}_{R} \ll H \rho_{R}\,. \label{eq:SR1}
\end{equation}
This regime can be parametrized by a set of slow-roll parameters $\epsilon$, $\eta$ and $\beta$,
defined by:\footnote{In some of the WI literature (including our own previous paper \cite{Montefalcone:2022jfw}), the slow-roll (SR) parameters are defined somewhat differently, namely as the quantities in the equation here divided by a factor $(1+Q)$. With these alternate definitions, the slow-roll conditions are satisfied instead for SR parameters $\ll1$. In other words, the factor $(1+Q)$ can be put on either side of the equation stating the SR condition.}
\begin{equation}
\epsilon=\frac{M^2_{\rm{pl}}}{2}\left(\frac{V_\phi}{V}\right)^2, \quad \eta=M^2_{\rm{pl}} \frac{V_{\phi \phi}}{V}, \quad \beta=M^2_{\rm{pl}}\left(\frac{\Upsilon_\phi V_\phi}{\Upsilon V}\right),
\end{equation}
which in order to meet the conditions in eq.~\eqref{eq:SR1}, must themselves satisfy: $\epsilon \ll 1+Q$, $|\eta| \ll 1+Q$ and $|\beta| \ll 1+Q$~\cite{Taylor:2000ze, Hall:2003zp, Bastero-Gil:2004oun, Moss:2008yb}. In other words, inflation quickly comes to an end when one of the slow-roll parameters defined above violates the slow-roll
condition, i.e.\ $\epsilon= 1+Q$, or $|\eta|=1+Q$ or $|\beta|=1+Q$.
More rigorously, the acceleration expansion continues as long as:
\begin{equation}
    \epsilon_H\equiv -\frac{\dot{H}}{H^2}\simeq \frac{\epsilon}{1+Q} <1, \label{eq:epsH}  
\end{equation}
and terminates when $\epsilon_H=1$.

To summarize, the background dynamics of WI obeys the system of three ordinary differential
equations described by eqs.~\eqref{eq:WI.1},~\eqref{eq:WI.2} and~\eqref{eq:WI.4}—together with two constraints in eqs.~\eqref{eq:Upsilon} and~\eqref{eq:WI.3}—which are evolved until $\epsilon_H=1$. We can rewrite this system of equations in terms of the number of $e$-folds:
\begin{equation}
    N_e\equiv\int H {\rm d}t, \label{eq:defNe}
\end{equation}
which quantifies the amount of expansion that the scale factor undergoes during the accelerated expansion.
Denoting derivatives
with respect to $N_e$ with the dagger symbol (not to be confused with Hermitian conjugation), the system of differential equations for the background evolution reads:
\begin{eqnarray}
    \phi^{\dagger\dagger}+3(1+Q)\phi^{\dagger} &=& -\frac{V_{,\phi}}{H^2}, \label{eq:BG1}\\
    \rho_R^{\dagger}+4\rho_R &=& 3H^2Q\phi^{\dagger 2},  \label{eq:BG2}\\
    H^2 &=& 2\frac{V(\phi)+\rho_R}{6M_{\rm{pl}}^2-\phi^{\dagger 2}}; \label{eq:BG3}
\end{eqnarray}
with $\dot{\phi} = H\phi^{\dagger}$, $\dot{\rho}_r= H\rho_r^{\dagger}$, and $\Ddot{\phi} \simeq H^2\phi^{\dagger\dagger}$ where we neglect slow-roll corrections.\footnote{We checked numerically that this assumption has no impact on the resulting background evolution.}

\section{Cosmological perturbations in warm inflation}
\label{sec:perturbations}

\subsection{Basic framework \label{sec:perturbations_1}}

The presence of a radiation bath and a dissipation rate not only alters the background dynamics of the inflaton field but also its perturbations. Specifically, thermal fluctuations of the radiation fluid source density perturbations that are subsequently transferred to the inflaton field as adiabatic curvature perturbations. This contrasts with standard cold inflation, in which the seeds for large-scale structure formation are simply generated by quantum fluctuations. To investigate scalar perturbations in WI, throughout this paper we follow the approach of~\cite{Ramos:2013nsa}. In this section we review the system of perturbation equations and solutions from that paper, although our notation differs in eqs.~\eqref{eq:3.14} to~\eqref{eq:Gab2.4} in that we choose an explicit parameterization with respect to $\tau=\ln(k/aH)$  for convenience.
Further, when we solve the system of equations, to simplify the numerical implementation in sections \ref{sec:code} and \ref{sec:results}, we set the dissipation strength $Q$ to a constant value during the inflationary expansion, instead of letting it evolve as in other work~\cite{Ramos:2013nsa}.\footnote{The robustness of this assumption is investigated thoroughly in section \ref{sec:results_1}.} In short, the system of perturbations' equations we investigate here is already well established. 

However, a comprehensive numerical examination of their solutions had not been performed to date. One of the scopes of our work is to fill this gap. Specifically, the key contribution of our paper, in addition to the public code presented in section~\ref{sec:code}, is the extensive study on the relationship between the scalar dissipation function $G(Q)$ in eq.~\eqref{eq:defG} and other model parameters. Further, our numerical analysis extends to higher values of $Q$ that what has been explored to date. All the aforementioned contributions are thoroughly presented in section~\ref{sec:results}. 

Returning now to the basic setup, we consider the perturbed FLRW metric in Newtonian gauge:\footnote{Although any gauge can be chosen, the Newtonian gauge is a simple and convenient choice that is numerically stable when integrating the full set of evolution equations~\cite{Das:2020xmh}.}
\begin{equation}
    {\rm d}s^2 = -(1 -2\varphi){\rm d}t^2 + a^2(t)(1+2\varphi)\delta_{ij}{\rm d}x^{i}{\rm d}x^{j}\,.
    \label{eq:2.1}
\end{equation}
We expand the inflaton field, radiation energy density, and radiation pressure around their background values:
\begin{align}
    \phi(\mathbf{x}, t) & =\bar{\phi}(t)+\delta \phi(\mathbf{x}, t)\,, \\
    \rho_r(\mathbf{x}, t) & =\bar{\rho}_r(t)+\delta \rho_r(\mathbf{x}, t)\,, \\
    p_r(\mathbf{x}, t) & =\bar{p}_r(t)+\delta p_r(\mathbf{x}, t)\,.
\end{align}
The evolution equations for the field and radiation perturbations follow from conservation of the energy-momentum tensor. In momentum space, the equations of motion for the radiation perturbations with comoving wavenumber $k$ read~\cite{Bastero-Gil:2011rva}:
\begin{align}
    & \delta \dot{\rho}_r+4 H \delta \rho_r=-3\dot{\varphi}\left(1+w_r\right) \rho_r +\frac{k^2}{a^2} \Psi_r+\delta Q_r-Q_r \varphi\,, \label{eq:2.2}\\
    &\dot{\Psi}_r+3 H \Psi_r+w_r \delta \rho_r=-\left(1+w_r\right) \rho_r \varphi+J_r\,, \label{eq:2.3}
\end{align}
where $\Psi_r$ is the momentum perturbation, $w_r\simeq 1/3$\footnote{Recall we are assuming $\rho_r\propto T^4$, i.e.\ that the system is close-to-equilibrium, as this is a necessary condition for the calculation of the dissipative coefficient to hold. Thus we are taking $p_r\simeq \rho_r/3 $ and, hence, $w_r = 1/3$.} is the equation of state for the radiation fluid, and $J_r = -\Upsilon\dot{\phi}\delta\phi$ and $Q_r = \Upsilon \dot{\phi}^2$ are the momentum and radiation source terms, respectively. The radiation source perturbation $\delta Q_r$ is given by~\cite{Bastero-Gil:2011rva}:
\begin{equation}
    \delta Q_r=\delta \Upsilon \dot{\phi}^2+2 \Upsilon \dot{\phi} \delta \dot{\phi}-2 \varphi \Upsilon \dot{\phi}^2, \label{eq:2.4}
\end{equation}
where
\begin{equation}
    \delta \Upsilon=\Upsilon\left[c \frac{\delta T}{T}-m \frac{\delta \phi}{\phi}\right]\,, \label{eq:2.5}
\end{equation}
according to its parametrization defined in eq.~\eqref{eq:Upsilon}.

In addition to eqs.~\eqref{eq:2.2} and~\eqref{eq:2.3}, we also have the evolution equation for the inflaton field perturbations $\delta\phi$. This is obtained by perturbing the inflaton field's equation of motion about its background value, and adding two stochastic Gaussian noise terms corresponding to the quantum and thermal fluctuations~\cite{Ramos:2013nsa}:
\begin{align}
& \delta \ddot{\phi}+3 H \delta \dot{\phi}+\left(\frac{k^2}{a^2}+V_{, \phi \phi}\right) \delta \phi+\delta \Upsilon \dot{\phi} +\dot{\phi}(3H\varphi+4\dot{\varphi}) \nonumber\\
&+(2 \ddot{\phi}+3 H \dot{\phi}) \varphi+\Upsilon(\delta \dot{\phi}+\varphi\dot{\phi})=\xi_q+\xi_T. \label{eq:2.6}
\end{align}
$\xi_T$ describes the thermal noise fluctuations in the local approximation for the effective equation of motion of the inflaton field. It is connected to the dissipation coefficient through a Markovian fluctuation-dissipation
relation, which in an expanding space-time is
given by~\cite{Ramos:2013nsa}:
\begin{equation}
    \langle \xi_{T}(\mathbf{k},t)\xi_{T}(\mathbf{k^\prime},t^\prime)\rangle=\frac{2\Upsilon T}{a^3} (2\pi)^3 \delta(\mathbf{k}-\mathbf{k}^\prime) \delta(t-t^\prime). \label{eq:2.7}
\end{equation}
On the other hand, $\xi_q$ describes the back-reaction of short-distance quantum modes on long-distance field modes as in the standard stochastic inflation formalism. $\xi_q$ can thus be interpreted as a Gaussian white noise term, with two-point correlation function~\cite{Kamali:2023lzq}:
\begin{equation}
    \langle \xi_{q}(\mathbf{k},t)\xi_{q}(\mathbf{k^\prime},t^\prime)\rangle=\frac{H^2 (9+12\pi Q)^{1/2}(1+2n_{*})}{\pi a^3} (2\pi)^3 \delta(\mathbf{k}-\mathbf{k}^\prime) \delta(t-t^\prime),  \label{eq:2.8}
\end{equation}
where $n_*$ denotes the inflaton statistical distribution due to the presence of the radiation bath. This is generally assumed to be the equilibrium Bose-Einstein distribution, i.e.\ $n_* = [\exp(H/T) - 1]^{-1}$. We emphasize that eq.~\eqref{eq:2.8} is an approximation of the quantum noise term's two-point correlation function. The quantum noise amplitude is chosen to reproduce the analytical estimate for the scalar power spectrum in WI derived in~\cite{Ramos:2013nsa}, under a rigorous treatment of the stochastic inflation approach to WI perturbations. We provide an explicit proof of this in section~\ref{sec:analytic_estimate}.

The complete system of first order perturbation equations for WI is obtained by combining eqs.~\eqref{eq:2.2}--\eqref{eq:2.6}. To solve this system of equations, we perform a convenient change of variables from cosmic time $t$ to: 
\begin{equation}
    \tau\equiv\ln z, \,\,\,\, {\rm{ where}} \,\,\,\, z\equiv k/(aH) \, . \label{eq:deftau}
    \end{equation}
    It can be easily shown that for any function $f(t)$ it follows:
\begin{align}
    \dot{f}  &= H(\epsilon_H-1)f^{\prime}\simeq -H f^{\prime} , \label{eq:chain_rule1}\\
    \ddot{f} &= H^2(\epsilon_H-1)^2 f^{\prime\prime}+H^2(\epsilon_H+\epsilon_H\eta_H-\epsilon_H^2)f^{\prime}\simeq H^2 f^{\prime\prime}, \label{eq:chain_rule2}
\end{align}
where $^\prime \equiv {\rm d}/{\rm d}\tau$, $\epsilon_H$ has been defined in eq.~\eqref{eq:epsH}, and $\eta_H\equiv H^{-1}{\rm d}\ln\epsilon_H/{\rm d}t$. Since the perturbations' equations are evolved until the CMB scale $k_{*}$ exits the horizon, approximately 60 $e$-folds before the end of inflation, both $\epsilon_H$ and $\eta_H$ are $\ll 1$  and can therefore be neglected.\footnote{We have checked this numerically as shown in figure~\ref{fig:5}.} Combining eqs.\ref{eq:defNe} and \ref{eq:deftau}, one notes that $\tau$  is related to the
number of $e$-folds by a simple linear transformation:
\begin{equation}
    N_e=-\tau+ \tau_k, \label{eq:Netotau}
\end{equation}
where $\tau_k$ represents an arbitrary additive constant that can be chosen to ensure that $\tau$ denotes the number of $e$-folds as measured before horizon crossing of the $k-$mode. 

In addition we also reparameterize all of the dynamical variables by $k^{3/2}$ and rescale the thermal 
 and quantum noise terms such that they have a two-point correlation function normalized to unity. In formulas: 
\begin{align}
&\{\delta\hat{\phi},\delta\hat{\rho}_r,\hat{\Psi}_r,\hat{\varphi}\}=k^{-3/2}\{\delta\phi,\delta\rho_r,\Psi_r,\varphi\}; \label{eq:3.14}\\
&\xi_{T,q}(\mathbf{x},t)\rightarrow \frac{\Xi_{\mathrm{T,q}}H^2e^{3\tau/2}}{k^{3/2}}\hat{\xi}_{T,q}(\mathbf{k},\tau),  \\
&\langle\hat{\xi}_{T,q}(\mathbf{k},\tau) \hat{\xi}_{T,q}(\mathbf{k}^\prime,\tau^\prime)\rangle=(2\pi)^3\delta\left(\tau-\tau^{\prime}\right) \delta^{(3)}(\mathbf{k}-\mathbf{k}^{\prime});
\end{align}
where,
\begin{equation}
    \Xi_{\mathrm{q}}\equiv H(9+12\pi Q)^{1/4}\sqrt{\frac{1+2n_*}{\pi}}, \qquad \Xi_T\equiv \sqrt{6H QT}. 
    \label{eq:noise_amps}
\end{equation}
In this notation, the complete system of perturbation equations reads:
\begin{align}
    &\delta \hat{\phi}^{\prime\prime}-3 (1+Q) \delta \hat{\phi}^\prime+\left(e^{2\tau}+\frac{V_{, \phi \phi}}{H^2}-\frac{3mQ \dot{\phi}}{\phi H}\right) \delta \hat{\phi} +\frac{ c}{4 H \dot{\phi}} \delta \hat{\rho}_r -4\frac{\dot{\phi}}{H}\hat{\varphi}^{\prime}  \nonumber \\
    &+\left[3H(1+2Q)\dot{\phi}+2\Ddot{\phi}\right]\frac{\hat{\varphi}}{H^2}=e^{3\tau/2}\sum_{j=\{q,T\}} \Xi_j \hat{\xi}_j\,, \label{eq:Gab2.1} \\
    &\delta \hat{\rho}^\prime_r+\left(\frac{3c Q \dot{\phi}^2}{4 \rho_r}-4\right) \delta \hat{\rho_r}+He^{2\tau}\hat{\Psi}_r-6 H Q \dot{\phi} \delta \hat{\phi}^\prime -\frac{3m Q \dot{\phi}^2}{\phi} \delta \hat{\phi}-3HQ\dot{\phi}^2\hat{\varphi}-4H\rho_r\hat{\varphi}^\prime=0, \label{eq:Gab2.2}\\ 
    &\hat{\Psi}^\prime_r-3 \hat{\Psi}_r-3Q \dot{\phi} \delta \hat{\phi}-\frac{1}{3H} \delta \hat{\rho}_r+4 \rho_r \frac{\hat{\varphi}}{3H}=0, \label{eq:Gab2.3} \\
      &\hat{\varphi}^\prime - \hat{\varphi}+ \frac{1}{2H M_{\mathrm{Pl}}^2}\left(\hat{\Psi}_r- \dot{\phi} \delta \hat{\phi}\right)=0. \label{eq:Gab2.4}
\end{align}
This system of equations can be solved numerically subject to the initial conditions represented by the vector $\{\delta \hat{\phi}_0,\delta\hat{\phi}^\prime_0,\delta \hat{\rho}^0_r, \hat{\Psi}^0_r,\hat{\varphi}_0\}$
and for a given background evolution for the quantities $\{H,Q,\rho_r,\phi,\dot{\phi},\ddot{\phi}\}$, according to the formalism described in section~\ref{sec:background_ev}.

The scalar power spectrum is determined from the definition of the comoving curvature perturbation $\mathcal{R}$, rescaled by $k^{3/2}$:
\begin{equation}
    \Delta^2_{\mathcal{R}}(k)=\frac{1}{2 \pi^2}\left\langle|\hat{\mathcal{R}}|^2\right\rangle, \label{eq:SPS}
\end{equation}
where $\langle \dots \rangle$ stands for the ensemble average over different realizations of the noise terms and:
\begin{align}
    \hat{\mathcal{R}}&=\sum_{i=\phi, r} \frac{\rho_i+\bar{p}_i}{\rho+\bar{p}} \hat{\mathcal{R}}_i, \\
    \hat{\mathcal{R}}_i&=-\hat{\varphi}-\frac{H}{\rho_i+\bar{p}_i} \hat{\Psi}_i,
\end{align}
with $\hat{\Psi}_\phi\equiv -\dot{\phi}\delta \hat{\phi}$, $\bar{p}=p_\phi+p_r$, $\bar{p}_\phi \equiv p_\phi=\frac{1}{2}\dot{\phi}^2-V$, $\rho_\phi=\frac{1}{2}\dot{\phi}^2+V$ and $\bar{p}_r\equiv p_r=\frac{1}{3}\rho_r$. 

Finally, it is noteworthy to emphasize that by opting for the time variable $\tau$ and rescaling all the dynamical variables by $k^{3/2}$, we effectively eliminate the $k$-dependence both in the perturbations’ equations (eqs.\ref{eq:Gab2.1} to \ref{eq:Gab2.4}) and in the equation for the resulting scalar power spectrum (eq.~\eqref{eq:SPS}). Consequently, this approach enables the evaluation of a universal, $k$-independent solution to the equations, while also providing a convenient means to reintroduce the $k$-dependence by setting $\tau = 0$, i.e. $k=aH$, at the time a specific mode $k$ of interest crosses the horizon. In our case for the CMB mode $k_{*}$, $\tau(k_*)=0$ corresponds to roughly 60 $e$-folds before the end of inflation.

Recently, ref.~\cite{Ballesteros:2023dno} presented an alternative formulation of the perturbations' equations (eqs.~3.2 to 3.7 in their manuscript) that  makes use of a Fokker-Planck approach to obtain a system of deterministic linear
differential equations for the two-point correlation function of the perturbations. This approach--previously introduced in ref.~\cite{Ballesteros:2022hjk}--differs from the methods employed in this work and consists of an interesting and novel pathway to solve the system of stochastic differential equations for the perturbations numerically. By employing this method, the need to repeatedly solve the complete system of stochastic differential equations for the perturbations is effectively circumvented.

In practice, a fair method for the assessment of the different approaches would consist e.g.\ in comparing the computational resources to achieve the desired accuracy. However, a discussion along this line for the Fokker-Planck approach has not been provided in ref.~\cite{Ballesteros:2023dno}. While we have detailed the specifics of our parallelized code in section~\ref{sec:perturbations_code}, we leave the study of the effectiveness of this formalism and its potential implementation in our publicly available code to future work.

In addition to the method employed, the formulation in ref.~\cite{Ballesteros:2023dno} departs from the one presented here in two ways: (1) the quantum noise term $\xi_q$ in the inflaton field perturbations is neglected, and (2) the thermal noise term $\xi_T$ has been incorporated in the radiation perturbations. Apart from these discrepancies, we have checked that the results from the two approaches agree as soon as the strong dissipative regime $Q > 1$ is considered, which is the relevant regime for the purposes of this work. Nevertheless, we briefly comment on these aspects.

With respect to point (1), in our framework, the quantum noise term  is necessary to accurately consider the impact of sub-horizon quantum fluctuations on the behavior of long wavelength modes in the inflaton field. By doing so, we achieve a comprehensive understanding of the combined effects arising from both inflaton quantum and thermal fluctuations, succinctly captured by a single stochastic Langevin-like equation of motion. Notably, in this formulation, the cold inflation limit ($Q=0$) emerges naturally, as eq.~\eqref{eq:Gab2.1} aligns with the conventional Starobinsky stochastic inflation approach~\cite{Starobinsky:1986fx} in this regime. In contrast, in ref.~\cite{Ballesteros:2023dno} the cold limit is attained from the homogeneous solution of the Langevin equation for the inflaton perturbations sourced by thermal noise alone. This is achieved under a specific choice of inflaton perturbations quantization, which allows for the separation of contributions to the power spectrum into an homogeneous and inhomogeneous component.  The two approaches exhibit inconsistency within the weak dissipative regime of warm inflation for $Q\ll 1$. Specifically, in ref.~\cite{Ballesteros:2023dno}, the scalar power spectrum generally appears smaller compared to our framework, seemingly underestimating the non-negligible contribution from the thermal fluctuations in the radiation bath and the thermal excitation of the inflaton perturbations for $T>H$.

Regarding point (2), the addition of a thermal noise term in the radiation perturbations had already been thoroughly studied in ref.~\cite{Bastero-Gil:2014jsa} proving that the thermal noise has only a minor influence on the resulting scalar power spectrum, which becomes negligible for a non-zero temperature power of the dissipation rate, i.e. $c\neq 0$.\footnote{We checked this aspect numerically and confirmed the results already presented in ref.~\cite{Bastero-Gil:2014jsa}.} Bastero-Gil {\it et al.} also emphasize in ref.~\cite{Bastero-Gil:2014jsa} the ambiguity of incorporating this term, which comes from the fact that Einstein equations do not fully fix whether the dissipative noise source enters in the energy or in the momentum flux. In conclusion, the decision to include or neglect the thermal noise term in the radiation perturbations does not undermine either approach, rather presents a relatively minor yet unresolved matter within the context of warm inflation that does not modify the outcome of this work.

\subsection{Analytic estimate of the power spectrum \label{sec:analytic_estimate}}

To obtain the scalar power spectrum, one  needs to solve the system of differential equations in eqs.~\eqref{eq:Gab2.1}--\eqref{eq:Gab2.4}. In general this can only be done numerically, since the radiation, inflaton and metric perturbations' equations are all mutually coupled. Specifically, from eqs.~\eqref{eq:Gab2.1} and~\eqref{eq:Gab2.2}, we see that the inflaton fluctuations couple directly to the radiation fluctuations when $c\neq 0$. As first shown in \cite{Graham:2009bf}, this coupling results in a growing (decaying) mode in the curvature power spectrum for $c>0$ ($<0$).

In this subsection, we will restrict our discussion to the one specific case in which  an explicit analytic expression for the scalar power spectrum can in fact be obtained: the case where  the temperature power of the dissipation rate is set to zero ($c=0$) and  the metric perturbations are set to zero by imposing $\varphi=0$ in the Newtonian gauge.\footnote{Recently, ref.~\cite{Ballesteros:2023dno} proposed analytical estimates in the case of a non-zero value of $c$, demonstrating its distinct and non-negligible influence on the power spectrum also in the weak dissipative regime, i.e.\ for $Q\ll 1$. Note, within the context of our work, that this is already captured by the scalar dissipation function $G(Q)$ as defined in eq.~\eqref{eq:GQ_definition}.} This second approximation is certainly valid for small to moderate values of the dissipation strength $Q$, as one can easily show that the coefficients of the $\varphi$ and $\varphi^\prime$ terms in the perturbations' equations are first order in the slow roll parameters $\epsilon$ and $\eta$. With these simplifications, the equations for the inflaton and the radiation perturbations decouple and one can obtain the curvature power spectrum using Green's function techniques. This was first done in \cite{Ramos:2013nsa}, in which the authors also presented a rigorous treatment of the quantum noise term using the stochastic inflation formalism. Below, we provide a simplified proof that reproduces the analytical estimate obtained in \cite{Ramos:2013nsa}, using the approximate Gaussian two-point correlation function for the quantum noise from eq.~\eqref{eq:2.8}, thus justifying its form.

Following the notation in ref.~\cite{Ramos:2013nsa}, the equation satisfied by the inflaton perturbations $\delta\phi$ in eq.~\eqref{eq:Gab2.1} can be written in terms of the variable $z\equiv k/(aH)$ as:
\begin{equation}
    \delta \phi^{\prime \prime}-\frac{1}{z}(3 Q+2) \delta \phi^{\prime}+\left[1+3 \frac{\eta-\beta Q /(1+Q)}{z^2 M_{\rm{pl}}^2}\right] \delta \phi =\frac{1}{H^2 z^2}\left[\xi_T+\xi_q\right]\,, \label{eq2:Gab5}
\end{equation}
which has general solution:
\begin{equation}
    \delta \phi(\mathbf{k}, z)=\int_z^{\infty} {\rm d} z^{\prime} \Phi\left(z, z^{\prime}\right) \frac{\left(z^{\prime}\right)^{1-2 \nu}}{z^{\prime 2} H^2}\left[\xi_q\left(z^{\prime}\right)+\xi_T\left(z^{\prime}\right)\right]\,. \label{eq:dphi1}
\end{equation}
Here, the Green's function $\Phi(z,z^\prime)$ is given in terms of Bessel functions:
\begin{equation}
    G\left(z, z^{\prime}\right)=\frac{\pi}{2} z^\nu z^{\prime \nu}\left[J_\alpha(z) Y_\alpha\left(z^{\prime}\right)-J_\alpha\left(z^{\prime}\right) Y_\alpha(z)\right]\,,
\end{equation}
which is solved within $z^\prime > z$ and where:
\begin{align}
    \nu & =3(1+Q) / 2, \label{eq:nu}\\
    \alpha & =\sqrt{\nu^2+3\left(\frac{ \beta
Q}{1+Q}- \eta\right)}\,. \label{eq:alpha}
\end{align}

Using Eq.~\eqref{eq:dphi1} and the definition of the inflaton power spectrum $\Delta^2_{\delta\phi}$ in terms of the two-point correlation function for the scalar field perturbations, we obtain:
\begin{align}
    \Delta^2_{\delta\phi}&\equiv \frac{k^3}{2 \pi^2} \int \frac{{\rm d}^3 k^{\prime}}{(2 \pi)^3}\left\langle\delta \varphi(\mathbf{k}, z) \delta \varphi\left(\mathbf{k}^{\prime}, z^\prime\right)\right\rangle \\
    &=\frac{k^3}{2 \pi^2} \int \frac{{\rm d}^3 k^{\prime}}{(2 \pi)^3} \frac{1}{H^4} \int_z^{\infty} {\rm d} z_2 \int_z^{\infty} {\rm d} z_1 G\left(z, z_1\right) G\left(z, z_2\right) \frac{\left(z_1\right)^{1-2 \nu}}{z_1^2} \frac{\left(z_2\right)^{1-2 \nu}}{z_2^2}\left[\left\langle\xi_q \xi_q\right\rangle +\left\langle\xi_T\xi_T\right\rangle \right] \nonumber \\
    &=\left[\frac{\Upsilon T}{16\pi^2}+\frac{H^2\sqrt{9+12\pi Q}}{ 32\pi^3}\cdot(1+2n_{*}) \right]\int_z^{\infty} {\rm d} z^{\prime} z^{\prime 2-4 \nu} G\left(z, z^{\prime}\right)^2. \label{eq:dphi2}
\end{align}
In the last line, we use the expressions for the two-point correlation functions of the thermal and quantum noises in eqs.~\eqref{eq:2.7} and~\eqref{eq:2.8}, and we assume that $\Upsilon$, $T$, $H$ and $Q$ are approximately constant with respect to $z$. Since $z\ll 1$ for the relevant observational scales for the inflaton perturbations, we can approximate the integral
appearing in eq.~\eqref{eq:dphi2} as:
\begin{align}
     \Delta^2_{\delta\phi}&\simeq\left[\frac{\Upsilon T}{16\pi^2}+\frac{H^2\sqrt{9+12\pi Q}}{ 32\pi^3}\cdot(1+2n_{\mathrm{*}}) \right] \frac{\left[2^\nu \Gamma(\nu)\right]^2 \Gamma(\nu-1)}{\Gamma(2 \nu-1 / 2) \Gamma(\nu-1 / 2)} \label{eq:gamma_expression} \\
 &\simeq \left[\frac{\Upsilon T}{16\pi^2}+\frac{H^2\sqrt{9+12\pi Q}}{ 32\pi^3}\cdot(1+2n_{\mathrm{*}}) \right]\frac{8\pi}{\sqrt{9+12\pi Q}} \\
 &=\frac{H^2}{4\pi^2}\left[1+2n_{\mathrm{*}}+\frac{T}{H}\frac{2\sqrt{3}\pi Q}{\sqrt{3+4\pi Q}} \right].
\end{align}
Here, $\Gamma(x)$ is the Gamma function. To obtain the final equality, we (1) use the asymptotic form for the Bessel function $Y_\alpha(z)$ when $z\ll 1$, and (2) neglect the terms proportional to the slow-roll coefficients, such that $\alpha\simeq \nu =3(1+Q)/2$.  

In a spatially flat gauge, the comoving curvature perturbation takes the simple form $\mathcal{R}= H\delta\phi/\dot{\phi}$~\cite{liddle_lyth_2000}. From this it follows:
\begin{align}
\Delta_{\mathcal{R},\rm{analytic}}^2&=\frac{H^2}{\dot{\phi}^2}\Delta_{\phi}^2=\left(\frac{H^2}{2\pi \dot{\phi}}\right)^2\left[1+2n_{\mathrm{*}}+\frac{T}{H}\frac{2\sqrt{3}\pi Q}{\sqrt{3+4\pi Q}} \right]\,.
    \label{eq:DeltaR_analytic}
\end{align}

In summary, the above analytical estimate for the scalar power spectrum in WI is only valid if the temperature power of the dissipation rate is to zero, i.e. $c = 0$, and if the
metric perturbations are neglected, i.e. setting $\varphi=0$ in the Newtonian gauge. In addition to these primary assumptions, to obtain the analytical expression in Eq.~\eqref{eq:DeltaR_analytic}, we had to additionally assume that: (1) the temperature $T$ and Hubble parameter $H$ are constant with respect to $z$ in order to bring them outside of the integral in eq.~\eqref{eq:dphi2}; (2) the term proportional to the slow-roll coefficient $\eta$ could be neglected as such that we could set $\alpha\approx\nu$. As we will show numerically in section \ref{section:results_analytic}, these two secondary assumptions, as well as neglecting the metric perturbations, do not significantly affect the resulting scalar power spectrum. In contrast, deviating from a temperature power of the dissipation rate equal to zero, i.e. $c\neq 0$, notably impacts the scalar power spectrum, particularly in the regime characterized by strong dissipation. This observation is what originally motivated the introduction of a multiplicative function of the dissipation strength $G(Q)$, in conjunction with the existing analytical expression. 

For general cases with $c \neq 0$, the analytic expression in eq.~\eqref{eq:DeltaR_analytic} must therefore be multiplied by a factor $G(Q)$
which can only be determined numerically.
The subsequent section delves into the definition and observational relevance of $G(Q)$, which we also refer to as the scalar dissipation function.

\subsection{The scalar dissipation function $G(Q)$ \label{sec:GQ_def}}

The analytic form of the power spectrum given by eq.~\eqref{eq:DeltaR_analytic} is the result used in most of the recent literature in WI. As emphasized above, once we include the coupling between the inflaton and
radiation perturbations, i.e. set $c\neq 0$, and restore the metric perturbations, these equations can only be solved numerically. This results in a correction to eq.~\eqref{eq:DeltaR_analytic} which is usually parametrized by a multiplicative function
of the dissipation strength $G(Q)$, determined by fitting the
numerically computed curvature power spectrum for a
given form of the dissipation rate. In formulas, this reads:
\begin{equation}
    G(Q)\equiv \frac{\Delta_{\mathcal{R},\rm{numerical}}^2|_{c\neq 0}}{\Delta_{\mathcal{R},\rm{analytic}}^2}. \label{eq:GQ_definition}
\end{equation}

Once the scalar dissipation function $G(Q)$ is specified, we can compute the spectral tilt $n_s$ and the tensor-to-scalar ratio $r$, and compare the obtained results with the observational constraints from the CMB at the pivot scale $k_*$: 
\begin{align}
    n_s&\simeq 1+\frac{{\rm d}\ln \Delta^2_{\mathcal{R}}}{{\rm d}N_e}= 0.9649\pm 0.0042,& \quad \text{at 68\% CL,~\cite{Planck:2018jri}} \label{eq:ns_Planck}, \\
   r &\equiv \frac{\Delta_{\mathcal{T}}^2}{\Delta^2_{\mathcal{R}}}\lesssim 0.036, & \quad \text{at 95\% CL.~\cite{BICEP:2021xfz}}. \label{eq:r_Planck}
\end{align}
Here, $\Delta_{\mathcal{T}}^2$ is the tensor power spectrum which in WI remains effectively unaltered relative to the standard cold inflationary scenario, since the gravitational interactions are weak.\footnote{In refs.~\cite{Bhattacharya:2006dm,Qiu:2021ytc} it was shown that a stimulated emission of gravitons in thermal equilibrium with the radiation bath  would require the temperature of the thermal bath to be $T \sim M_{\rm pl}$, which is much higher than what is achieved during WI. For other effects that potentially alter the scalar and tensor spectra, see refs.~\cite{Namba:2015gja,Peloso:2016gqs}. }  Thus, the curvature power spectrum is the quantity primarily affected by the dissipation in WI, which in turn significantly alters the resulting spectral tilt $n_s$ and tensor-to-scalar ratio $r$, relative to standard cold inflation. In order to test WI models against current data, it is therefore  crucial to obtain precise numerical fits of $G(Q)$ for different forms of the dissipation rate and check their robustness across different inflaton potentials, model parameters and inflationary histories. This constitutes the main subject of this paper and of the publicly available code that we describe in detail in the following section.

\section{Code implementation \label{sec:code}}

The bulk of the code is contained in the Python script \texttt{WI\_Solver\_utils.py}, which  is structured in four modules: (1) \texttt{Inflaton\_Model}, (2) \texttt{Background}, (3) \texttt{Perturbations} and (4) \texttt{Scalar\_Dissipation\_function}, which respectively (1) define the main parameters for a given inflaton potential, solve for its (2) background and (3) perturbations' evolution, and finally use this information to (4) fit for the scalar dissipation function $G(Q)$. An important simplification we make throughout this code is to take $Q$ as a constant during the inflationary expansion. In section~\ref{sec:results_1} we show explicitly that this assumption has no impact on the obtained fits for $G(Q)$, while greatly simplifying the numerical implementation.

e\subsection{\texttt{Inflaton\_Model}}
\label{sec:Inflaton_Model}

In the \texttt{Inflaton\_Model} module, we define the inflaton potential and consequently all of the parameters that depend on it (such as the Hubble parameter $H$, its derivative with respect to cosmic time $\Dot{H}$, etc.), that will be used for all the subsequent computations for the background and perturbations' evolution. Several potentials have been proposed to embed the warm inflationary dynamics into a larger framework. Here, we review the ones currently available in the code, which generally represent some of the most studied cases in the literature.

{\bf Monomial potential.} The simplest type of potential we consider is the power-law form:
\begin{equation}
    \label{eq:monomial_potential}
    V(\phi)= \frac{\lambda}{n!}\,\phi^n\,,
\end{equation}
where the index $n$ describes the steepness of the potential and $\lambda$ is dimensionful except when $n=4$, for which it describes the self-coupling of the inflaton field. The form in eq.~\eqref{eq:monomial_potential} has been adopted as the inflationary potential since the early conceptions of the theory, although it is now strongly disfavored by CMB data in combination with BAO and BICEP2/Keck Array data in the context of the standard cold inflation~\cite{Planck:2018jri}. In contrast, in WI, due to the different dynamics resulted
from the dissipative effects, monomial potentials, specifically with even  power law $n$, are consistent with CMB observations for a
large range of the parameter values~\cite{Benetti:2016jhf}.

{\bf Hilltop-like potential.} One of the simplest alternatives to monomial potentials is of the form:
\begin{equation}
    V(\phi)=V_0\left[1-\left(\frac{\phi}{\phi_{\rm{f}}}\right)^{2n}\right]^{2}
    \label{eq:hilltop}
\end{equation}
with $n\geq 1$ and with the inflaton rolling off the top (plateau) of a potential, i.e.\ starting at $\phi\ll \phi_{\rm f}$~\cite{Boubekeur:2005zm}.  We assume  $\phi_{\rm f}$ to be sufficiently large such that inflation ends before the inflection point of the potential is reached. For a potential of the form in eq.~\eqref{eq:hilltop}, the field excursion during inflation is much smaller than the Planck scale, which is generally preferred from an effective field theory perspective. Hilltop-like potentials were found to be consistent with CMB data in both cold and warm inflation for wide regions of parameter space~\cite{Bartrum:2013oka,Benetti:2016jhf}.

{\bf Natural inflation.}  The model of Natural Inflation~\cite{Freese:1990rb} uses axion-like particles as natural candidates for the inflaton, in that their potentials possess an underlying shift symmetry that protects the required flatness of the potential to achieve both sufficient inflation and the correct amplitude of density perturbations for structure formation~\cite{Adams:1990pn,Montefalcone:2022owy}.  The original variant of Natural Inflation used a cosine potential  motivated by the case of the QCD axion:

\begin{equation}
    \label{eq:natural_potential}
    V(\phi)= \Lambda^4\,\left(1+\cos(N\phi/f)\right)\,,
\end{equation}
where $\Lambda$ is related to the magnitude of the explicit breaking scale and $N$ is the number of vacua. The model avoids the $\eta$-problem intrinsic to cold inflation and introduces a nearly flat potential suitable for slow-roll inflation. These symmetry properties also suppress radiative and thermal corrections which allow this model and more generally axion-like potentials to be trivially implemented in the context of WI~\cite{Visinelli:2011jy, Mishra:2011vh,Kamali:2019ppi,Berghaus:2019whh}. This potential in the WI scenario was found to be consistent with observations for super-Planckian and marginal sub-Planckian values for the
decay constant $f$~\cite{Correa:2022ngq,Montefalcone:2022jfw}.

{\bf $\beta$-exponential potential.} A generalization of power-law (runaway) inflation models is the $\beta$-exponential potential of the type~\cite{Alcaniz:2006nu}:
\begin{equation}
    \label{eq:betaexponential}
    V(\phi)= V_0\,\left[1-\lambda\,\beta\,\frac{\phi}{M_{\rm pl}}\right]^{1/\beta}\,,
\end{equation}
which arises from theories of gravity with higher derivative terms or compactified Kaluza-Klein models. Here, $\lambda$ is a dimensionless coupling and the constant $\beta$ parametrizes the deviation from the pure exponential model (for $\beta\to0$). In cold inflation scenarios, CMB data favors a non-minimal gravitational coupling of the inflaton in the case of $\beta$-exponential potentials~\cite{dosSantos:2021vis}, while these models are consistent with WI with a cubic dissipation rate~\cite{Santos:2022exm}.

\subsection{\texttt{Background}
\label{sec:background}}

The \texttt{Background} module solves for the background evolution as a function of the number of $e$-folds $N_e$ according to the system of equations in eqs.~\eqref{eq:BG1}--\eqref{eq:BG3}, for a given choice of the inflaton potential and its parameters. The solver evolves the system of equations from $N_e=0$ to $N_e=N^{\rm{end}}_{e}\equiv N_e^{\rm{pert}}+N_e^{\rm{infl}}$, with the initial conditions chosen such that $\epsilon_H=1$, i.e.\ inflation ends, when $N_e=N_e^{\rm{end}}$.

$N_e^{\rm{infl}}$ refers to the number of $e$-folds between horizon crossing of the CMB mode $k_*$ and the end of inflation. We generally take this to be $60$ to match CMB observations.
$N_e^{\rm{pert}}$ refers to how long before horizon crossing is an individual mode initialized when evolving the perturbations. This sets the duration of numerical evolution of a particular mode---we take this value to be approximately 5 to 10 $e$-folds. Note that this choice does not necessarily initialize the perturbations at the start of the background solution, as doing so can lead to numerical stability issues. We elaborate on this in Section \ref{sec:perturbations_code} below.\footnote{
To be more precise, the perturbation equations are only evolved for a specified interval in $\tau \in[\tau_{\rm{max}},\tau_{\rm{min}}]$ that contains $\tau=0$, i.e.\ the time corresponding to CMB horizon crossing. The number of $e$-folds $N_e^{\rm{pert}}$ is then safely chosen such that it contains said $\tau$ range up to CMB horizon crossing, i.e.\ $N^{\rm{pert}}_e \supseteq[\tau_{\rm{max}},0]$. See section~\ref{sec:perturbations_code} for more details.
}

To find the value of $\phi_0$ that guarantees $N_e^{\rm{end}}$ $e$-folds before inflation ends, we define the function \texttt{Bg\_solver\_test} which is an iterative solver of the background evolution that, starting from an initial guess $\phi_0^{\rm{guess}}$, scans the parameter space in $\phi$ via a standard gradient descent algorithm until the condition $\epsilon_H(N_e^{\rm{end}})=1$ is satisfied; we refer to Appendix~\ref{Appendix:ICs} for more details.
 
For given values of $Q$ and $\phi_0$, the additional required initial conditions on $\rho_{r,0}$ and $\phi^\dagger_0$ are determined from eqs.~\eqref{eq:BG1}--\eqref{eq:BG3} assuming slow-roll, i.e.\ for $\phi^{\dagger\dagger}_0=0$ and $\rho^{\dagger}_0=0$. In formulas:
\begin{align}
    & \rho_{r,0}=-\frac{Q V_{,\phi}(\phi_0)}{4(1+Q)}, \quad
-\frac{V_{,\phi}(\phi_0)}{3(1+Q)\phi_0^\dagger}=2\frac{V(\phi_0)+\rho_{r,0}}{6M_{\rm{pl}}^2-\phi^{\dagger 2}_0}.
\end{align}

Once the initial conditions are determined, we use the function \texttt{Bg\_solver\_tau} which additionally outputs all the relevant background quantities for the perturbations' equations as a function of $\tau \equiv \ln(k/(aH))$, such that they can be input directly in the perturbations solver.\footnote{
Note that solving the background with respect to the number of $e$-folds $N_e$ is a convenient choice since as shown in eq.~\eqref{eq:Netotau}, $N_e$ is linearly related to the time variable $\tau$. This implies that the background evolution with respect to $N_e$ and $\tau$ is the same up to a linear transformation of the independent variable, i.e.\ the background solution can be directly inputted in the perturbations' equations.
}

We emphasize here that the simplifying assumption of a constant dissipation strength $Q$ during the inflationary expansion implies that the background evolution is independent of both the temperature and inflaton field's power-law exponents in the dissipation rate $\Upsilon$, respectively $c$ and $-m$ according to eq.~\eqref{eq:Upsilon}. This feature of the code is crucial as it allows for an efficient evaluation of the scalar dissipation function $G(Q)$, as we will show in greater detail in sec.~\ref{sec:results}. However, it also implies that currently \texttt{WarmSPy} should not be directly employed to study the background dynamics in WI, where the explicit evolution of $Q$ is essential. Note, that the effectiveness of our code resides in the modularity and the capability to solve for the inflaton perturbations given various backgrounds for the evolution of the inflaton field. In any case, in future updates of \texttt{WarmSPy}, we plan to enhance the code's capabilities by incorporating the explicit evolution of $Q$ and enable a precise study of the background dynamics in WI.

\subsection{\texttt{Perturbations}}
\label{sec:perturbations_code}

The \texttt{Perturbations} module takes as input (1) the background evolution computed in the \texttt{Background} module for a given choice of the inflaton potential, and (2) the values for $c$ and $m$, respectively the temperature and negative inflaton field's power-law exponents in the dissipation rate $\Upsilon$. The module numerically solves the perturbations' stochastic equations of motion, \eqref{eq:Gab2.1}--\eqref{eq:Gab2.4}, for a specified interval in $\tau\in [\tau_{\rm{max}},\tau_{\rm{min}}]$ divided into $N_\tau$ steps. This interval is chosen to include $\tau=0$, the time corresponding to horizon crossing of the mode being evolved. Since in this study we are interested in the CMB mode $k_{*}$, we specifically set $\tau=0$ at the time corresponding to $N_e^{\rm{inf}}$ $e$-folds before the end of inflation, i.e. when the CMB mode $k_{*}$ crosses the horizon, in formulas this corresponds to:\footnote{Recall that we evaluate the background evolution in the interval $N_e\in[0,N_e^{\rm{end}}]$ with $N_e^{\rm{end}}\equiv N_e^{\rm{pert}}+N_e^{\rm{infl}}$. This means that the time of CMB horizon crossing, i.e. $N_e^{\rm{inf}}$ $e$-folds before the end of inflation, corresponds to $N_e=N_e^{\rm{end}}-N_e^{\rm{infl}}=N_e^{\rm{pert}}$. In other words, $\tau(k_*)=0$ when $N_e=N_e^{\rm{pert}}$.}
\begin{equation}
    \tau(N_e^{\rm{pert}})=0.
\end{equation} 

The system of equations is then solved over $N_\xi$ realizations of the stochastic noises and from that the scalar power spectrum at $\tau=0$ is computed according to its definition in eq.~\eqref{eq:SPS},  i.e.\ as the ensemble average of the comoving curvature perturbation $\mathcal{R}$. The default $\tau$ interval is given by $\{\tau_{\rm{max}},\tau_{\rm{min}}\}=\{6.5,-1\}$. Additionally, we choose $N_\tau=5\times10^{5}$ time steps and $N_\xi=1024$ realizations of the noise terms. The value of $\tau_{\rm{min}}$ is set so that all modes are evolved for 1 $e$-fold after horizon crossing. On the other hand, the choices on $\tau_{\rm{max}}$, $N_\tau$ and $N_\xi$ were made in order to ensure numerical stability in the solution, without compromising the overall speed of the code. 

In this regard, the selection of $\tau_{\rm{max}}$ plays a crucial role in preserving the analytic accuracy and numerical stability of our solution, as it is closely tied to the initialization of the $k$-mode evolution. As noted in Section \ref{sec:background}, we do not necessarily initialize the mode at the start of inflation. Rather, we start the mode's evolution at some fixed time between 5 and 10 $e$-folds before horizon crossing. This choice becomes particularly important for scenarios where inflation lasts a considerable duration, such as $\mathcal{O}(1000)$ $e$-folds. In such cases, the term $e^{2\tau}$ in the equation of motion \eqref{eq:Gab2.1} becomes extremely large at early times. Recalling that $\tau$ denotes the number of $e$-folds as measured before horizon crossing (see eq.~\eqref{eq:Netotau}), this yields an $\mathcal{O}(e^{2000})$ term in eq.~\eqref{eq:Gab2.1} when the mode is initialized. The level of precision required to resolve both these early-time large terms and the other smaller terms that are relevant at late times would dramatically slow down any numerical evolution of the perturbations' equations.

To overcome this issue, we exploit the fact that most of the subhorizon evolution is trivial, since the mode will merely behave as a Minkowski harmonic oscillator in the deep subhorizon~\cite{Weinberg:2008zzc}. If we initialize the mode sufficiently early that it still falls in the Minkowski oscillator limit, our choice of initial conditions will still be accurate and applicable. This typically holds around 5 or more $e$-folds before horizon crossing, which eliminates the need to begin the mode's numerical evolution at the start of inflation. In the above example of inflation lasting $\mathcal{O}(1000)$ $e$-folds, instead of the aforementioned $\mathcal{O}(e^{2000})$ blow-up in the equations of motion, this simplification only has a term of $\mathcal{O}(e^{10})$, which is much more manageable numerically.
With our default choice for $\tau_{\rm{max}}$, the evolution of the perturbations' equations is initialized 6.5 $e$-folds before horizon exit, which recovers the subhorizon oscillator limit just discussed. In section~\ref{sec:results_2}, we investigate the robustness of the obtained $G(Q)$ fits for different choices of $\tau_{\rm{max}}$ in greater detail.

As already discussed in section~\ref{sec:perturbations_1}, to solve for the evolution of the perturbations we also need to specify a set of initial conditions, which for simplicity are taken to be:
\begin{equation}
  \{\delta\hat{\phi}_0,\delta\hat{\rho}_{r,0},\hat{\Psi}_{r,0},\hat{\varphi}_0\}=\{\Xi_{T}^{1/2}(\tau_{\rm{max}}),0,0,0 \}.  
\end{equation}
where $\Xi_{T,0}^{1/2}$ is determined via eq.~\eqref{eq:noise_amps}, for a given value of the dissipation strength $Q$ and the corresponding background quantities $H(\tau_{\rm{max}})$ and $T(\tau_{\rm{max}})$ computed in the \texttt{Background} module. We emphasize here that the choice of initial conditions is actually not relevant. In fact, the stochastic sources dominate the evolution of the perturbations, washing out any information regarding the initial state of the perturbations~\cite{Bastero-Gil:2011rva,Ballesteros:2022hjk}.

This module also provides the option to turn on or off (1) the metric perturbations, i.e.\ setting $\varphi\neq0$ or $=0$, and (2) the corrections of first and second order in $\epsilon_H$ and $\eta_H$ from the chain rule when changing variables from cosmic time $t$ to $\tau$; see eqs.~\eqref{eq:chain_rule1} and~\eqref{eq:chain_rule2}. By default, we neglect both effects, since, as we will show explicitly in section~\ref{sec:results_2}, including them merely slows down the execution of the code without impacting the obtained fits for $G(Q)$.

Finally, the evaluation of the stochastic differential equations is completely parallelized via the function \texttt{Pool\_Solver}, over the number of available cores $N_{\rm cores}$ in a given machine. This significantly reduces the total execution time of the code roughly by a factor of $N_{\rm cores}$. For a machine with $N_{\rm{cores}}=128$ and given the default $\tau$ interval with $N_{\tau}=5\times10^{5}$, it takes roughly $20$ seconds to solve the perturbations' equations $N_\xi=1024$ times.

To summarize, the \texttt{Perturbations} module outputs the scalar power spectrum at CMB horizon crossing and its standard deviation as a function of the dissipation strength $Q$, which in this work we take to be in the interval $Q\in [10^{-7},10^{4}]$.

\subsection{\texttt{Scalar\_Dissipation\_function}}

The \texttt{Scalar\_Dissipation\_function} module takes the background and perturbations' evolution as inputs, computed respectively in the \texttt{Background} and \texttt{Perturbations} modules for a given choice of the inflaton potential and values of $c$ and $m$. The scalar dissipation function $G(Q)$ is then computed according to its definition in eq.~\eqref{eq:GQ_definition}. This module then determines the best fitting function $G(Q)$ via the method of least squares. Specifically, to allow for a direct comparison with previous work, we employ the polynomial fitting functions most prominently used in the literature due to its simplicity~\cite{Bastero-Gil:2016qru,Benetti:2016jhf,Montefalcone:2022jfw, Motaharfar:2018zyb}:
\begin{equation}
G_{\rm{pol}}(Q)=\begin{cases}
1+AQ^{\alpha}+BQ^{\beta},  & c>0 \\
(1+AQ^{\alpha})(1+BQ^{\beta})^{-\gamma}, & c<0
\end{cases}\label{eq:GQ_pol} 
\end{equation}
In the case of an inverse temperature
dependence in the dissipation rate ($c=-1$), $G(Q)_{\rm{pol}}$ fits our data well. On the other hand, for $c>0$, this fitting function does not robustly capture the behaviour of the scalar dissipation function over the entire range of values of $Q$ investigated, i.e. for $Q\in [10^{-7}, 10^{4}]$. To correct for this, when $c>0$ we propose a logarithmic fitting function of the form:
\begin{equation}
    G_{\rm{log}}(Q)=10^{\sum_{n=1}^4 a_n x^n}, \quad x\equiv\log_{10}.(1+Q)\label{eq:GQ_log}
\end{equation}
As we show in detail in section~\ref{sec:results}, $G_{\rm{log}}(Q)$ fits the data well over the entire range of values of $Q$ investigated, for the same number of free parameters as the polynomial fit, totaling four parameters.

\section{Results \label{sec:results}}

In this section, we present the numerical results, obtained with \texttt{WarmSPy}, for the scalar dissipation function $G(Q)$. We provide precise fits of $G(Q)$ for different values of the power-law temperature
dependence in the dissipation rate, namely $c\in \{3, 1, -1\}$, and investigate the relationship of the dissipation function $G(Q)$
to additional model parameters, such as the inflaton potential and the inflaton field power-law dependence of the dissipation rate. Unless otherwise stated, all results presented below were obtained with all the main parameters of the code set to their default values, specifically $\{\tau_{\rm{max}},\tau_{\rm{min}}\}=\{6.5,-1\}$, $N_\tau=5\times10^{5}$, $N_\xi=1024$, $N_e^{\rm{pert}}=10$ and $\varphi=0$.\footnote{See section~\ref{sec:perturbations_code} for the definition of these parameters and the motivation for said default values.}

\subsection{Method validation and comparison with previous literature
\label{sec:results_1}}

\begin{figure}[ht!]
	\centering
    \includegraphics[width=1\linewidth]{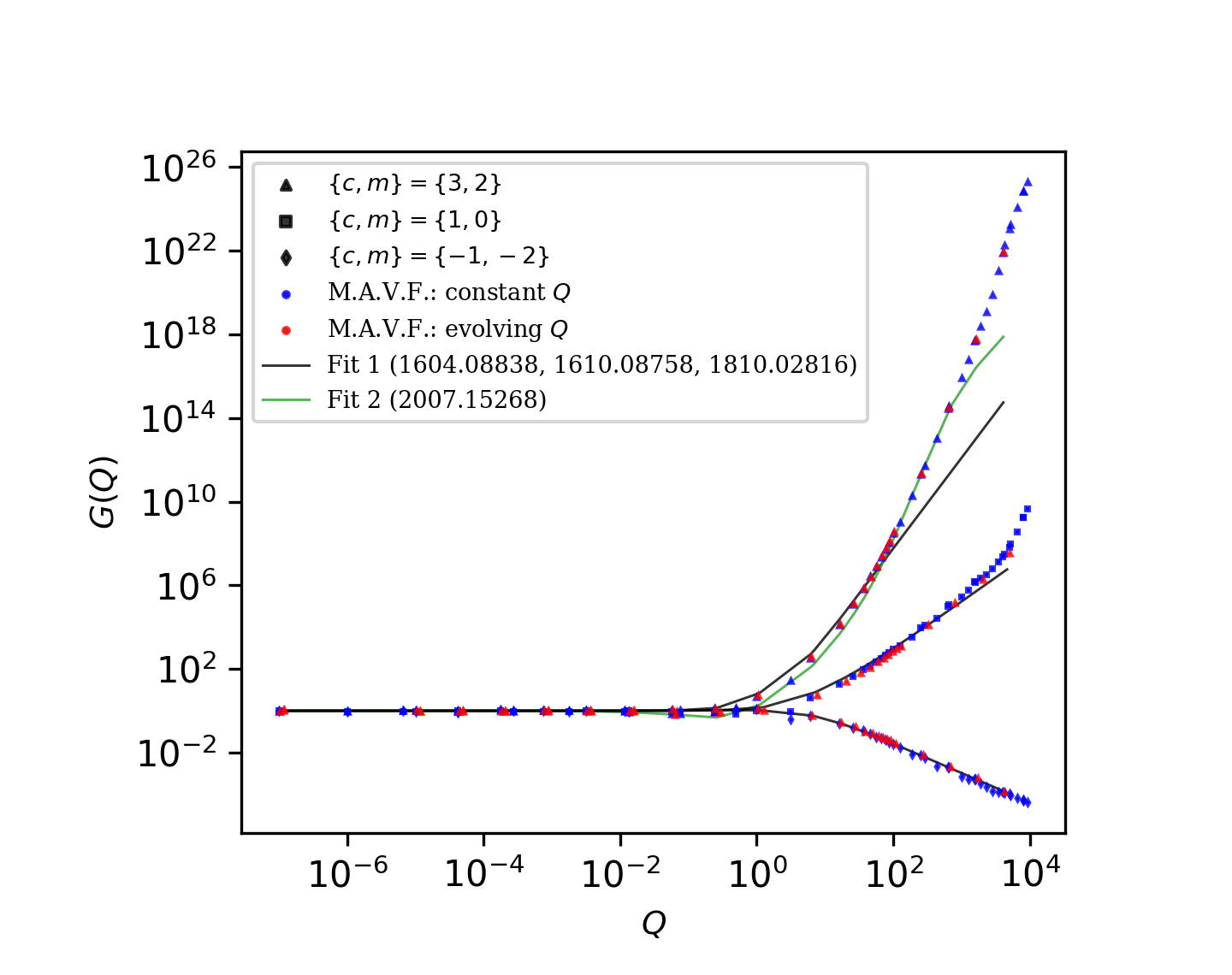}
\caption{The scalar dissipation function $G(Q)$ appearing in eq.~\eqref{eq:GQ_definition} as a function of the dissipation strength $Q \equiv \Gamma/(3H)$, for different values of the set of indices $(c,m)$ describing the power-law dependence of the dissipation rate to $T$ and $\phi$ respectively, see eq.~\eqref{eq:Upsilon}. Triangle: $(c,m) = (3,2)$; square: $(c,m) = (1,0)$; diamond: $(c,m) = (-1,-2)$. Results are shown both for the code employing a constant value of $Q$ (blue markers) and for an evolving $Q$ (red markers). Also shown are the fitting functions obtained from refs.~\cite{Bastero-Gil:2016qru,Benetti:2016jhf,Motaharfar:2018zyb} (labeled ``Fit 1'' in the legend) and in ref.~\cite{Das:2020xmh} (labelled ``Fit 2'' in the legend), see text for additional details.  In summary, this figure demonstrates the consistency of our code in the resulting $G(Q)$ when either keeping $Q$ constant or allowing it to evolve, as well as its agreement with the well-established $G(Q)$ fits found in the existing literature.}
	\label{fig:1}
\end{figure}

As mentioned in section~\ref{sec:code}, 
the evolution of the dissipation strength $Q$ is neglected during the inflationary expansion.
Such an approximation is one of the main 
differences
in the methodology between our work and the previous literature. Our choice greatly simplifies the evaluation of the initial conditions to obtain $N^{\rm{tot}}_e$ $e$-folds of inflation while keeping a high level of accuracy in the numerical solutions as discussed in the following sections. In fact, if we let $Q$ vary according to eq.~\eqref{eq:Upsilon}, the background evolution becomes dependent on the exponents $c$ and $m$. This increases the total number of computations required to obtain the scalar power spectrum for different combinations of $c$ and $m$, as one needs to redetermine the initial conditions each time. On the other hand, with our method, for a given inflaton potential and value of $N^{\rm{tot}}_e$, we only need to find the initial conditions once, regardless of the values of $c$ and $m$. 

Physically, taking $Q$ to be constant is equivalent to enforcing a specific evolution of the dimensionless constant $C_\Upsilon$ in the dissipation rate $\Upsilon$, see eq.~\eqref{eq:Upsilon}, in such a way as to keep the same dissipation strength $Q$ throughout the inflationary expansion. 
However, we would expect that once the numerical scalar spectrum is divided by the analytical result,
any potential alteration to the shape of the scalar power spectrum would get washed out, effectively leading to the same $G(Q)$. 
We confirm this assumption numerically by 
comparing
the solution found with a constant $Q$ against 
what is obtained by allowing $Q$ to vary according to eq.~\eqref{eq:Upsilon}. 

In figure~\ref{fig:1}, we show the comparison between the scalar dissipation function $G(Q)$ obtained with our code (M.A.V.F.\ in the legend) both for a constant and varying dissipation strength $Q$, and for different combinations of $c$ and $m$ (see the caption of figure~\ref{fig:1} for more details). 
Results are obtained assuming that the inflaton rolls down the quartic potential in eq.~\eqref{eq:monomial_potential} with $n=4$ and the self-interacting constant $\lambda=10^{-14}$, and setting $N_e^{\rm{infl}} = 60$.

In addition, figure~\ref{fig:1} shows the fitting functions $G(Q)$ obtained in refs.~\cite{Bastero-Gil:2016qru,Benetti:2016jhf,Motaharfar:2018zyb} (labelled ``Fit 1'' in the legend) where the cases $c=\{3,1,-1\}$ are explored respectively, along with the analytic form given in eq.~\eqref{eq:GQ_pol}. These fitting functions were also obtained for a quartic inflaton potential in refs.~\cite{Bastero-Gil:2016qru,Benetti:2016jhf,Motaharfar:2018zyb}, but for a smaller range of values $Q \lesssim 100$. As expected, the data produced with our code match these fitting functions extremely well for $Q<100$ and only diverge significantly for larger values of the dissipation strength and for $c=3$. Finally, we also plot a more recent fitting function computed in ref.~\cite{Das:2020xmh} (labelled ``Fit 2'' in the legend) for the case of a runaway potential, a cubic temperature dependence $c=3$ in the dissipation rate, and a wider range of values $Q \lesssim 10^{3}$. This fitting function reproduces our data more accurately than the basic polynomial form, and does so up to  $Q \sim 10^{3}$. This was not necessarily expected given that our data and the fitting function from ref.~\cite{Das:2020xmh} were obtained for different inflaton potentials, and it hints at the universality of the scalar dissipation function $G(Q)$ for a given value of $c$, an aspect we explore more in depth in the following section. Finally, it is also important to emphasize that the better fit achieved by the scalar dissipation function $G(Q)$ proposed in ref.~\cite{Das:2020xmh} is also due to its cumbersome form with 11 free parameters compared to the 4 from the simpler polynomial fit in eq.~\eqref{eq:GQ_pol}.

Overall, the main results from the comparison shown in figure~\ref{fig:1} can be summarized as follows: (1) there is no difference in the resulting scalar dissipation function $G(Q)$ when $Q$ is kept constant or left evolving during the inflationary expansion; (2) the scalar dissipation function $G(Q)$ evaluated with our code reproduces accurately the well-known $G(Q)$ fits previously obtained in the literature.

This discussion also raises two pertinent issues for further investigation: (1) the feasibility of generating precise fits of the scalar dissipation function $G(Q)$ for a large range of $Q$ values, i.e.\ for $Q\in[10^{-7},10^{4}]$, with a minimal number of free parameters; (2) the relationship (if any) of the scalar dissipation function $G(Q)$ for a set value of $c$ to additional model parameters, such as the inflaton potential or the value of $m$. These aspects will be thoroughly examined in the next section. 

\subsection{Numerical fits of the scalar dissipation function $G(Q)$ \label{sec:results_fits}}

\begin{figure}[ht!]
	\centering
    \includegraphics[width=1\linewidth]{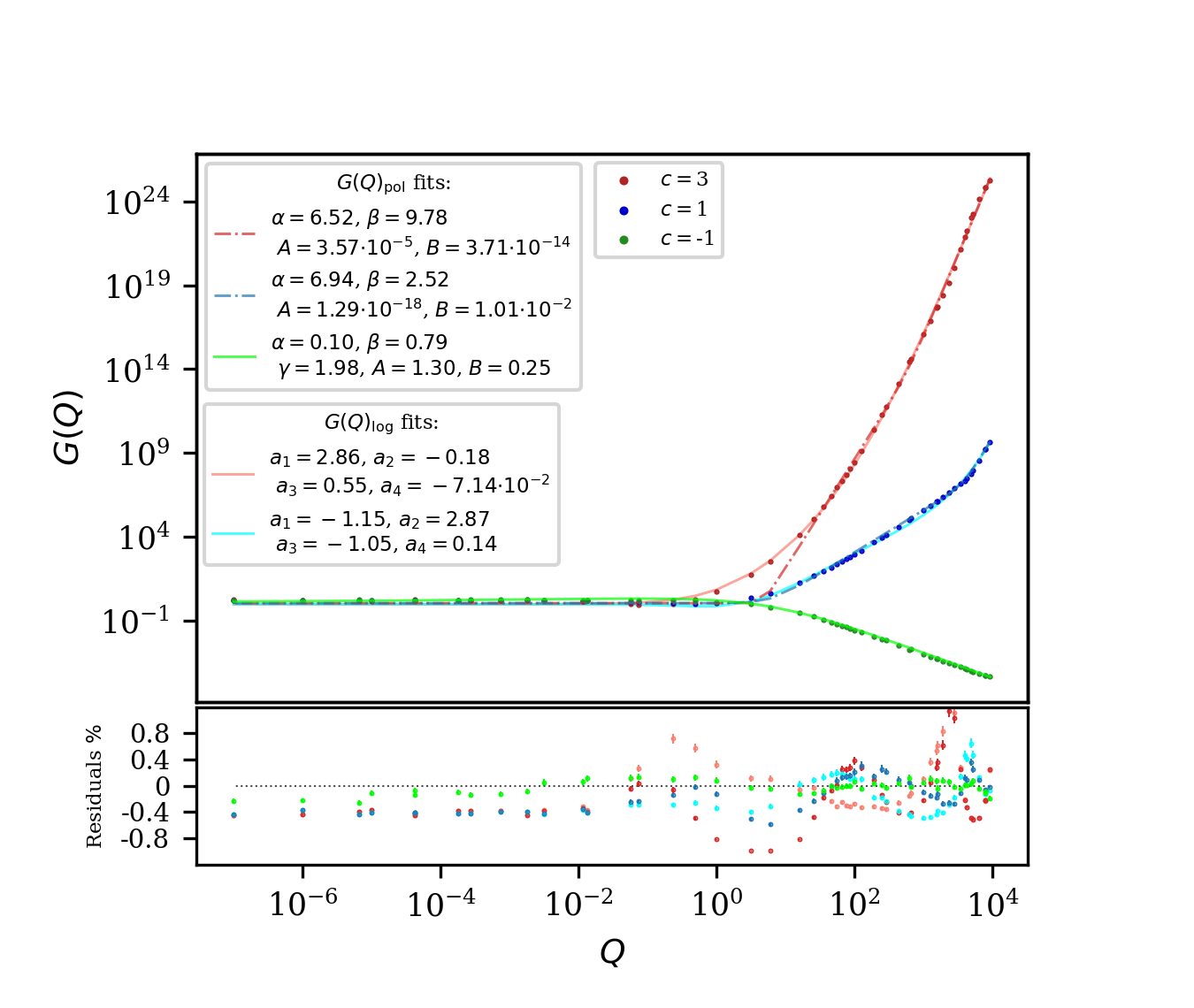}
	\caption{{\it Upper panel:} The numerical solutions obtained with our code for the scalar dissipation function $G(Q)$ as a function of $Q$, along with an assessment of the goodness of the fitting functions in eqs.~\eqref{eq:GQ_pol} and \eqref{eq:GQ_log} with the fitting parameters reported in the caption. We have considered a monomial potential, see eq.~\eqref{eq:monomial_potential}, with $n = 4$ and $\lambda = 10^{-14}$. Results are shown for a different temperature dependence in the dissipation rate (see eq.~\eqref{eq:Upsilon}): $c = 3$ (red), $c = 1$ (blue), $c = -1$ (green). {\it Lower panel:} The fractional residuals between the numerical result and the two different choices of the fit for each value of $c$, over the range of the dissipation strength $Q$ considered.}
	\label{fig:GQ_fits}
\end{figure}

In figure~\ref{fig:GQ_fits}, we plot the numerical solutions obtained with \texttt{WarmSPy}, along with the corresponding fits of the scalar dissipation function $G(Q)$ for different values of the power-law temperature dependence in the dissipation rate (see eq.~\eqref{eq:Upsilon}), namely $c\in\{3,1,-1\}$.

\begin{table}[ht!]
	\centering
	\begin{tabular}{c|ccccc}
		\toprule
		\hline 
		\multicolumn{6}{c}{$G_{\rm{pol}(Q)}$ }\\
		\hline \diagbox[width=15em]{Value of $c$}{Fit parameters} & $A$ & $B$ & $\alpha$ & $\beta$ &  $\gamma$ \\
		\hline
		3 & $3.57\times10^{-5}$ & $3.71\times10^{-14}$ & 6.52 & 9.78 & - \\
		1 & $1.23\times10^{-2}$ & $7.37\times10^{-17}$ & 2.48 & 6.49 & - \\
		-1 & 1.30 & 0.25 & 0.10 & 0.79 & 1.98 \\
		\hline
		\multicolumn{6}{c}{$G_{\rm{log}}(Q)$}\\
  \hline
  & $a_1$ & $a_2$ & $a_3$ & $a_4$ & \\
		 \hline
		3 & 2.86 & -0.18 & 0.55 & $-7.14\cdot 10^{-2}$ & - \\
		1 & -1.15 & 2.87 & -1.05 & 0.14 & - \\		
		\hline
		\bottomrule
	\end{tabular}
	\caption{The parameters obtained from the fitting functions in eqs.~\eqref{eq:GQ_pol} and \eqref{eq:GQ_log} to the numerical results for $G(Q)$ for shown in figure~\ref{fig:GQ_fits}.}
	\label{tab:1}
\end{table}
These results were obtained under the following assumptions: (1) there is no inflaton field dependence in the dissipation rate, i.e.\ $m=0$; (2) the inflaton rolls down the monomial potential in eq.~\eqref{eq:monomial_potential} with $n=4$ and the self-interacting constant $\lambda=10^{-14}$; (3) the number of $e$-folds after CMB horizon crossing is set to  $N_e^{\rm{infl}}=60$; (4) the metric perturbations are neglected, i.e.\ $\varphi=0$. 
While these assumptions might appear excessively specific and restrictive, the detailed analysis presented in section~\ref{sec:results_2} will demonstrate that the fits of the scalar dissipation function $G(Q)$ shown here possess broader validity, extending beyond the fulfillment of all aforementioned assumptions. In short, we find that the behaviour of the scalar dissipation function $G(Q)$ is susceptible only to the temperature dependence of the dissipation rate $c$. For the case of $c=\{3,1\}$ ($c=-1$), this behaviour is accurately described by the fitting function from eq.~\eqref{eq:GQ_log} (eq.~\eqref{eq:GQ_pol}) respectively for the parameters shown in table~\ref{tab:1}, throughout the entire range of values of $Q$ considered. Specifically for $c=\{3,1\}$, the fractional residuals plotted in the lower panel of figure~\ref{fig:GQ_fits} are always $\lesssim\mathcal{O}(1)$, meanwhile for $c=-1$ the fit is even more robust with fractional residuals $\lesssim \mathcal{O}(0.2)$.

For a positive temperature dependence in the dissipation rate, we also test the polynomial fitting function from eq.~\eqref{eq:GQ_pol} which is generally unable to capture the behavior of the scalar dissipation function $G(Q)$ over the entire range of $Q$ values considered. While this is not evident from the fractional residuals shown in the bottom panel of figure~\ref{fig:GQ_fits} which are roughly the same order as those for the logarithmic fit; one can note that for both $c=\{3,1\}$: (1) the best-fit for the parameter $B$ is driven to an extreme small value $\sim \{10^{-14}, 10^{-17}\}$ respectively; (2) the fit largely underestimates the numerical solution in the range $Q\in (1,10^{2})$. This seems to be at least partially related to the small dip in the magnitude of $G(Q)$ for $Q\sim \mathcal{O}(0.1-1)$, which cannot be captured by the simple polynomial fitting function. In addition, we also observed that the best fit parameters and the accuracy of the polynomial fit depends strongly on the range of values of $Q$ to which it is applied. This explains the significant difference in the best-fit parameters found here against those obtained in previous work, despite the consistency of the numerical solutions in the overlapping range of values of $Q$, see figure~\ref{fig:1}.  Given these considerations and acknowledging the value of a simple fitting function for $G(Q)$ in calculating cosmological observables, we suggest restricting the evaluation of the polynomial fit to a narrow range in $Q$ that is relevant to the specific inflaton model under examination. 

\subsection{Universality of the scalar dissipation function $G(Q)$ \label{sec:results_2}}

In this section, we assess the relationship of the scalar dissipation function $G(Q)$ for a set value of the temperature power in the dissipation rate $c\in \{3,1,-1\}$, to additional model parameters. Specifically, in the order in which they will be presented, we investigate the dependence of $G(Q)$ from: (1) the negative inflaton field power of the dissipation rate $m$; (2) the inflaton potential; (3) the number of $e$-folds before and after CMB horizon crossing, i.e.\ by respectively changing $\tau_{\rm{max}}$ and $N_e^{\rm infl}$. Finally we also analyze (4) the impact of turning on  the metric perturbations, i.e.\ setting $\varphi\neq0$, and/or the corrections of first and second order in $\epsilon_H$ and $\eta_H$ (see differential chain rule in eqs.~\eqref{eq:chain_rule1} and~\eqref{eq:chain_rule2}).
\begin{figure}[ht!]
    \centering
    \includegraphics[width=\linewidth]{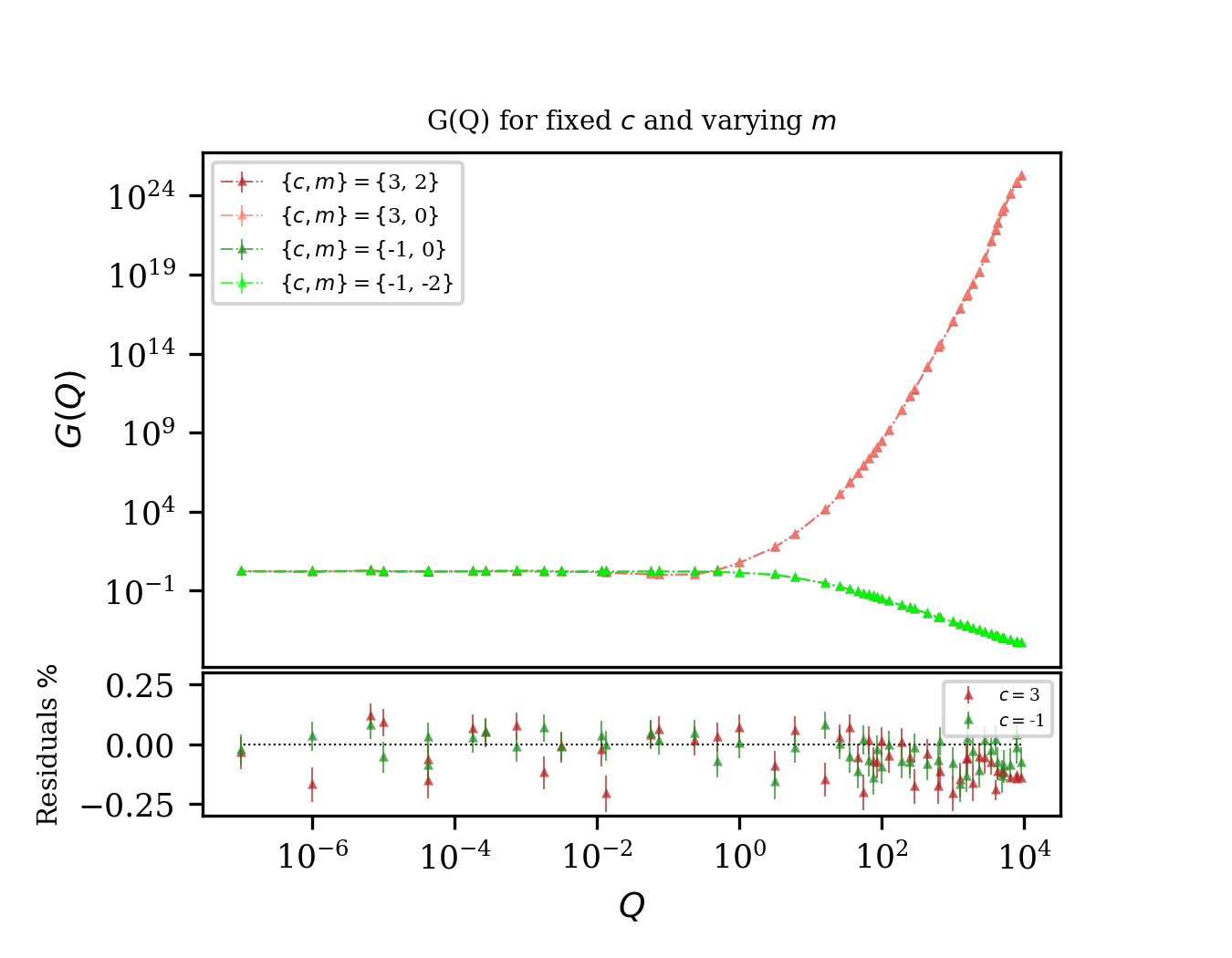}
    \caption{{\it Upper panel:} An assessment of the impact of the index $m$ in Eq.~\eqref{eq:Upsilon} on the numerical results of the scalar dissipation function $G(Q)$, for different temperature dependence in the dissipation rate, namely $c = 3$ (red), $c = -1$ (green). {\it Lower panel:} The fractional residuals between the different choices of the index $m$ over the range of the dissipation strength $Q$ considered. In summary, this figure demonstrates that $G(Q)$ is independent on the value of $m$.}
    \label{fig:2}
\end{figure}

\paragraph{The negative inflaton field power of the dissipation rate $m$:}
We discuss the impact of considering different values for the index $m$ (see eq.~\eqref{eq:Upsilon} for its definition) for a set temperature dependence in the dissipation rate.
Specifically, for this study we assume that the inflaton rolls down the monomial potential in eq.~\eqref{eq:monomial_potential} with $n=4$ and the self-interacting constant $\lambda=10^{-14}$, and setting $N_e^{\rm{infl}} = 60$. In the upper panel of figure~\ref{fig:2} we then plot the resulting scalar dissipation function $G(Q)$ for a cubic ($c=3$) and inverse ($c=-1$) temperature dependence of the dissipation rate and different values of $m$, respectively $m=\{2,0\}$ and $m=\{-2,0\}$.\footnote{Note, these values of $m$ are motivated by the several forms of dissipation rates derived from first principles in quantum field theory in the context of WI~\cite{Kamali:2023lzq}. For examples of these explicit constructions see \cite{Berera:1998gx,Berera:1998px,Bastero-Gil:2012akf,Berghaus:2019whh}.} For a given value of $c$, the scalar dissipation function $G(Q)$ is independent of $m$. Specifically, the lower panel of figure~\ref{fig:2} shows the residuals between the two values of $m$ assessed for each set value of $c=\{3,-1\}$, which all lie below $ \mathcal{O}(0.25)$ for the entire range of dissipation strength $Q$ considered. Additionally, one can note that, while still small, the residuals for the case with $m\neq0=c-1$ against the $m=0$ case are mostly negative as $Q\gtrsim 10^{2}$. This implies that the case with $m=0$ slightly underestimates the growth (suppression) of the scalar power spectrum in the strong dissipative regime when $c=3$ ($c=-1$).

These observations can be understood analytically by the fact that the $m$ dependence in the inflaton field perturbation equation is slow-roll suppressed in eq.~\eqref{eq:Gab2.1}, i.e.\ $m\dot{\phi} Q/(\phi H)\simeq \beta Q/(1+Q)$ which is negligible for $Q\lesssim 1$. In the strong dissipative regime ($Q\gg 1$), this term cannot be ignored and contributes to the growth (suppression) of the inflaton field perturbations for $c=\{3,-1\}$ respectively. Interestingly, even in this regime, this term has only a minor influence in the resulting scalar power spectrum as the coefficient proportional to $c$ multiplying $\delta\hat\rho_r$ dominates the overall evolution of the inflaton perturbations.

\begin{figure}[ht!]
    \centering
    \vspace{-1cm}
    \includegraphics[width=0.78\linewidth]{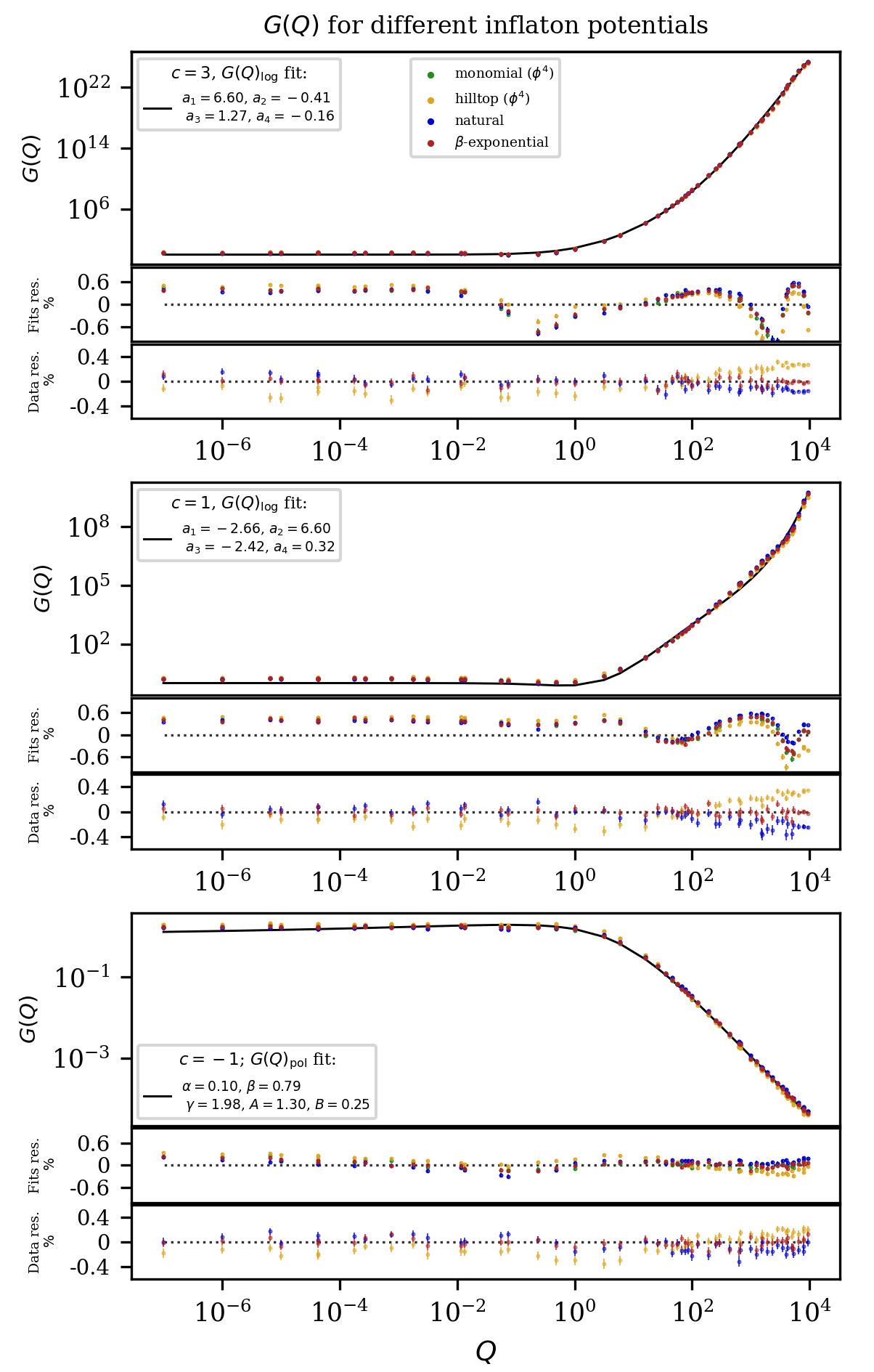}
    \caption{The results of our numerical code for the various inflaton potentials discussed in section~\ref{sec:background} and for a different temperature dependence in the dissipation rate: $c = 3$ (upper panel), $c = 1$ (middle panel),  $c = -1$ (lower panel). Details for the potentials used are in the main text. For each panel, we show the numerical results of  $G(Q)$ for the various potentials considered, the fractional residuals with respect to the best fits for $G(Q)$ from table~\ref{tab:1} and the fractional residuals with respect to the numerical solution for the monomial potential. In summary, this figure demonstrates that $G(Q)$ is independent on the inflaton potential chosen.}
    \label{fig:3}
\end{figure}

\paragraph{The inflaton potential:}
We discuss the impact of considering different models for the inflaton potential, following the framework discussed in section~\ref{sec:Inflaton_Model}. For this study, we assume $m=0$ in eq.~\eqref{eq:Upsilon}, i.e.\ no field dependence on the dissipation rate, set $N_e^{\rm{infl}}=60$, and compare the resulting scalar dissipation function $G(Q)$ for a given temperature dependence in the dissipation rate $c\in\{3,1,-1\}$ and different choices of the inflaton potential, namely the monomial, hilltop-like, natural and $\beta-$exponential  potentials. More precisely, we assume $n=4$ and $\lambda=10^{-14}$ for the monomial potential; $n=2$ and $\phi_{\rm{f}}=10\,M_{\mathrm{pl}}$ for the hilltop-like potential; $\Lambda=10^{16}\,$GeV, $f=5\,M_{\mathrm{pl}}$ and $N=1$ for the natural inflation potential; $V_0=10^{-23}\,M_{\mathrm{pl}}^4$, $\lambda=0.05$ and $\beta=0.3$ for the $\beta$-exponential potential.

The results obtained are presented in figure~\ref{fig:3} which is divided in 3 self-contained sub-figures respectively for the cases $c=\{3,1,-1\}$. Each sub-figure is in turn composed of an upper panel with the numerical solutions of $G(Q)$ for the different inflaton potentials considered, and two lower panels showing the fractional residuals respectively in reference to the best fits for $G(Q)$ from table~\ref{tab:1} and to the numerical solution for the monomial potential.

For a given value of $c$, the scalar dissipation function $G(Q)$ is independent of the inflaton potential. Specifically, the two lower panels of each sub-figure in figure~\ref{fig:3} show that the fractional residuals in reference to the best fits for $G(Q)$ and to the numerical solution for a monomial potential are respectively always $\lesssim \mathcal{O}(1)$ and $\lesssim\mathcal{O}(0.3)$, over the entire range of dissipation strength $Q$ considered. In addition, one can note that the shape of the residuals with respect to the best fits for $G(Q)$ reported in table~\ref{tab:1} (and obtained using a monomial potential) is the same regardless of the inflaton potential considered, further emphasizing that such fits are universally valid and independent of the inflaton model implemented. To the best of our knowledge, this observation is shown here explicitly for the first time. The implication is very powerful: one can compute with our code the dissipation function $G(Q)$ for a convenient choice of the inflaton potential, such as the monomial case, and apply it to any, perhaps more complex, inflaton model of interest.

In any case, one may still wonder if the results just presented depend on the specific values chosen for the parameters of the different inflaton potentials.  The choice was conveniently made in order to guarantee that inflation can last for a sufficient number of $e$-folds and also roughly reproduce the amplitude of the scalar power spectrum measured by CMB data, i.e.\ $\Delta^2_{\mathcal{R}}(k_*)\simeq 2.1\times10^{-9}$ with $k_*=0.05{\rm\,Mpc}^{-1}$~\cite{Planck:2018vyg}. Nevertheless, for a given inflaton potential the resulting scalar dissipation function is independent of the inflaton parameters chosen. This is expected as the main effect of a different choice of parameters is in the overall scale of the scalar power spectrum, which gets integrated out in order to obtain $G(Q)$. For the interested reader, we refer to figure~\ref{fig:A1} in Appendix \ref{Appendix:plots} which shows this explicitly for the monomial potential.

\begin{figure}[ht!]
    \centering
    \includegraphics[width=0.9\linewidth]{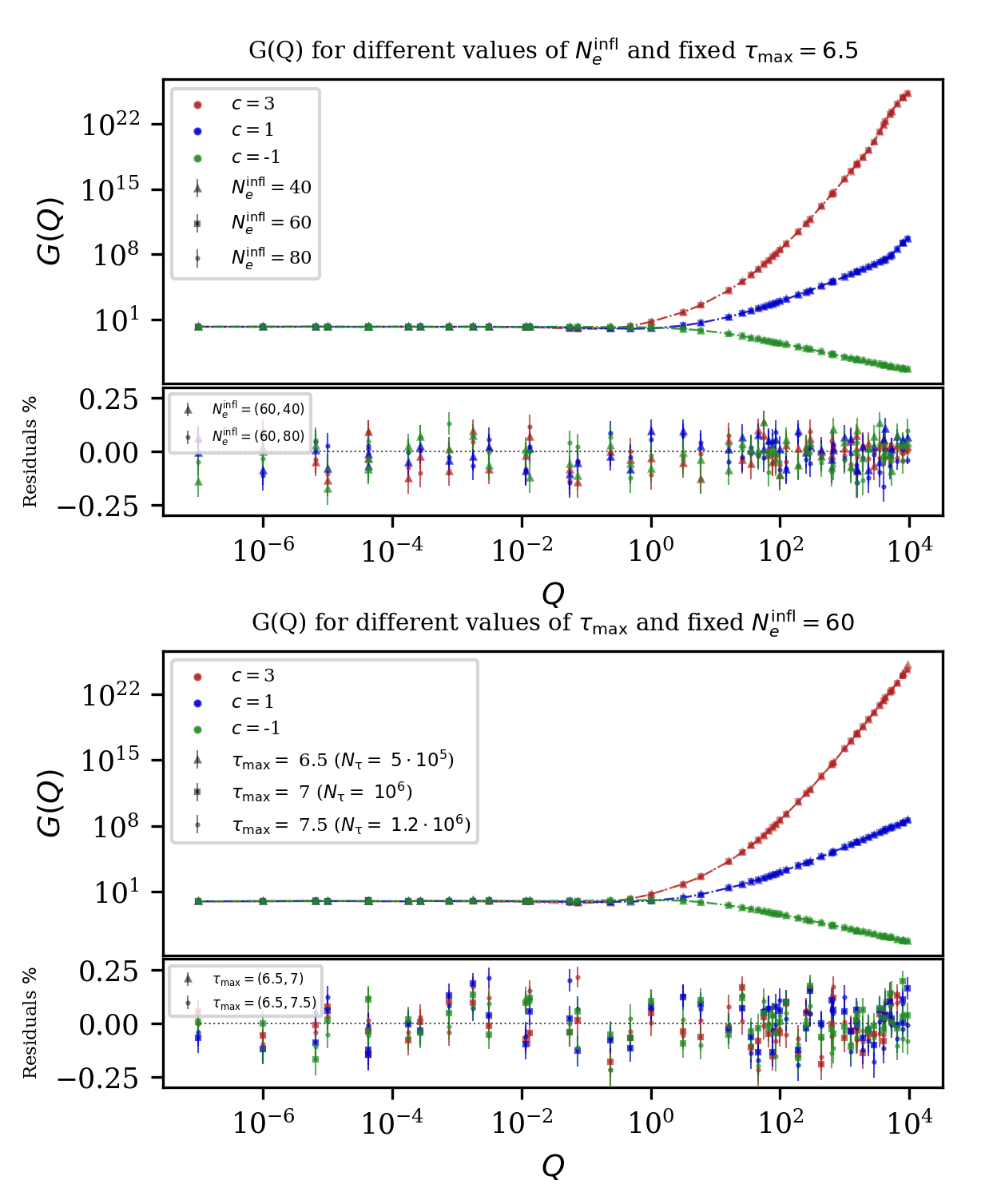}
\caption{ The scalar dissipation function $G(Q)$ obtained for different choices of the number of $e-$folds before and after CMB horizon crossing, respectively changing $\tau_{\rm{max}}$ (lower panel) and $N_e^{\rm{infl}}$ (upper panel). Details on the values of $\tau_{\rm{max}}$ and $N_e^{\rm{infl}}$ can be found in the text. In each panel, we show the  numerical results for $G(Q)$ for different temperature dependence in the dissipation rate,  $c = 3$ (red), $c = 1$ (blue),  $c = -1$ (green), and the corresponding fractional residuals with respect to the fiducial case with $N_e^{\rm{infl}}=60$ and $\tau_{\rm{max}}=6.5$. In summary, this figure demonstrates that $G(Q)$ is robust against changes in the inflationary history.}
    \label{fig:4}
\end{figure}

\paragraph{Number of $e$-folds before and after CMB horizon crossing:}
We assess the changes due to a variation in the number of $e$-folds after CMB horizon crossing $N_{e}^{\rm infl}$ and the quantity $\tau_{\rm{max}}$ that parametrizes the duration of the inflation stage considered before CMB horizon crossing. For this study, we assume no field dependence on the dissipation rate, i.e.\ $m=0$, and we consider the monomial potential in eq.~\eqref{eq:monomial_potential} with $n=4$ and the self-interacting coupling constant $\lambda=10^{-14}$. The resulting scalar dissipation function $G(Q)$ is compared for a given temperature dependence in the dissipation rate $c\in \{3, 1, -1\}$ and different choices of the number of $e$-folds before and after CMB horizon crossing, respectively changing $\tau_{\rm{max}}$ and $N_{e}^{\rm infl}$. Specifically, we present two main scenarios in the upper and lower plot of figure~\ref{fig:4}: (1) we fix $\tau_{\rm{max}}=6.5$ and vary the number of $e$-folds after CMB horizon crossing with $N_{e}^{\rm infl}=\{40,60,80\}$; (2) we fix $N_{e}^{\rm infl}=60$ and vary the number of $e$-folds before CMB horizon crossing over which we solve the perturbations' equations by setting $\tau_{\rm{max}}=\{6.5,7,7.5\}$.

For a given value of $c$, the scalar dissipation function $G(Q)$ is independent of the values for $N_e^{\rm infl}$ and $\tau_{\rm{max}}$. Specifically, the two lower panels in figure~\ref{fig:4} for each main plot show the fractional residuals between the different values of $N_e^{\rm{infl}}$ and $\tau_{\rm{max}}$ assessed for the range of indexes $c=\{3,1,-1\}$ considered, respectively. The residuals all lie below $\mathcal{O}(0.25)$ for the entire range of dissipation strength $Q$ considered. 

With respect to both scenarios analyzed, a change in the number of $e$-folds before or after CMB horizon crossing effectively results in a modification of the evolution of the background quantities and/or the initial conditions in the perturbations' equations. As previously mentioned in section~\ref{sec:perturbations_code}, changing the initial conditions has no appreciable impact on the solution to the perturbations' equations because the stochastic sources end up dominating the evolution, washing out any information regarding the initial state. On the other hand, changing the evolution of the background quantities in the perturbations' equations affects the resulting power spectrum, as one would expect. This evidently gets erased when the power spectrum resulting from the numerical computation is divided by the analytical estimate of the power spectrum to yield the scalar dissipation function $G(Q)$. In other words, for a given value of $Q$, the different evolution of the background quantities only alters the scale of the scalar power spectrum  which is then removed when evaluating $G(Q)$.

Finally, it is useful to clarify here the relationship between the length of the $\tau$ interval $N_{\rm{\tau}}$, the chosen value of $\tau_{\rm{max}}$ and the dissipation strength $Q$, in terms of how they impact the numerical stability of the solution of the perturbations' equations. Generally, as one would expect, the larger the value of $\tau_{\rm{max}}$, the larger the value for $N_{\rm{\tau}}$ is required. The relationship between $\tau_{\rm{max}}$ and $N_{\rm{\tau}}$ is highly non-linear such that choosing $\tau_{\rm{max}}\simeq 8$ requires setting $N_{\rm{\tau}}\gtrsim 10^7$ which significantly slows down the code execution time. This makes such choice for $\tau_{\rm{max}}$ undesirable. In addition, we note two more relevant aspects: (1) for a given $\tau_{\rm{max}}$, the smaller the value of $Q$, the larger $N_{\tau}$ needs to be to fully capture the numerical solution of the perturbations; (2) a larger value of $Q$ requires a larger value of $\tau_{\rm{max}}$ to allow the numerical solution to stabilize to its classical ensemble-average result. These considerations emphasize that varying $N_{\tau}$ and $\tau_{\rm{max}}$ in terms of the dissipation strength $Q$ would be the optimal choice from a computational standpoint, particularly if we want to extend the evaluation of the perturbations to values of the dissipation strength $Q$ above $\sim 10^{5}$. For simplicity, in this work we choose a fixed value for $\tau_{\rm{max}}$ and $N_{\tau}$ over the entire range of dissipation strength $Q$ considered, which we have checked numerically to guarantee a stable solution.

\paragraph{Metric perturbations and  corrections in $\epsilon_H$ and $\eta_H$:} 

\begin{figure}[ht!]
    \centering
    \includegraphics[width=\linewidth]{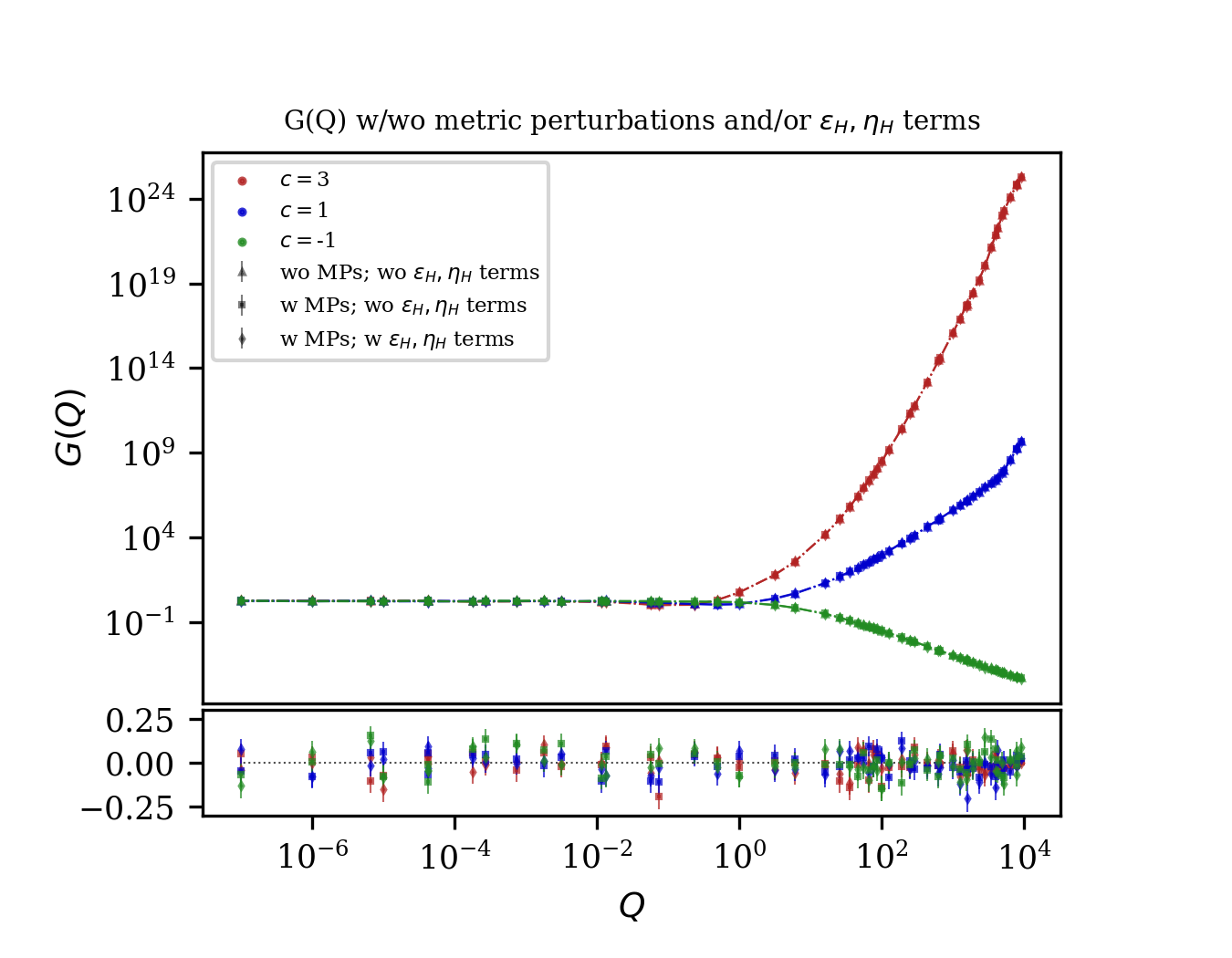}
    \caption{{\it Upper panel:} The scalar dissipation function $G(Q)$ obtained from considering different sources of corrections to the inflaton perturbations: no corrections (triangle), corrections from metric perturbations (square), corrections from both metric perturbations and slow-roll parameters (diamond). Results are shown for the index $c = 3$ (red), $c = 1$ (blue),  $c = -1$ (green). {\it Lower panel:} The fractional residuals between the results obtained with no corrections against i) setting only MPs different from zero or ii) both corrections from MPs and slow-roll parameters included in the computation. In summary, this figure demonstrates that $G(Q)$ is independent on the metric perturbations and slow-roll corrections.} \label{fig:5}
\end{figure}

We now assess the impact of the metric perturbations (MPs) $\varphi$ in eq.~\eqref{eq:2.1} and the corrections from the slow-roll parameters $\epsilon_H$ and $\eta_H$ on the numerical results. In fact, as discussed in section~\ref{sec:perturbations_code}, both MPs and the slow-roll corrections to the dynamics of the perturbations have been neglected in the previous analyses, however they can be reinserted through a switch that turns them back on in the numerical solution.
For this study, we assume no field dependence on the dissipation rate, i.e.\ $m=0$, and we consider the dynamics obtained from the monomial potential in eq.~\eqref{eq:monomial_potential} with the index $n=4$ and the self-interacting coupling constant $\lambda=10^{-14}$. We additionally set $N_e^{\rm infl}=60$. In the upper panel of figure~\ref{fig:5} we compare the resulting scalar dissipation function $G(Q)$ for a given temperature dependence in the dissipation rate $c\in\{3,1,-1\}$ without the metric perturbations and the corrections in $\epsilon_H$ and $\eta_H$ (triangle), with the corrections from metric perturbations included in the computation (square) or with both corrections from MPs and the slow-roll parameters included (diamond).

Neither the metric perturbations nor the corrections in the slow-roll parameters leads to a significant impact on the scalar dissipation function $G(Q)$ for the values of the index $c$ considered. More specifically, the lower panel of figure~\ref{fig:5} shows the fractional residuals between the case with $\varphi=0$ and no $\epsilon_H$ and $\eta_H$ corrections against the two other cases where $\varphi\neq 0$ and the $\epsilon_H$ and $\eta_H$ corrections are included or not, see the caption of figure~\ref{fig:5} for additional details. For all the values of $c$ assessed and in the entire range of dissipation strength $Q$ considered, the fractional residuals are $\lesssim\mathcal{O}(0.25)$ and centered around zero.

In regards to the corrections from the slow-roll parameters, this result was trivially expected since the equations for the perturbations are evolved up to the time of CMB horizon crossing, that is $N_e^{\rm{infl}}=60$ $e$-folds before the end of inflation, when both $\epsilon_H$ and $\eta_H$ are safely $\ll 1$  and can therefore be safely neglected. On the other hand, concerning the metric perturbations, the coefficients of the $\varphi$ and $\varphi^\prime$ terms in the perturbations' equations are first order in the slow roll parameters $\epsilon$ and $\eta$ and are thus negligible in the weak dissipative regime for $Q<1$. In the strong dissipative regime ($Q\gg 1$), while these terms should not be neglected they evidently have no impact in the resulting scalar power spectrum as the coefficients proportional to $c$ multiplying $\delta\hat{\rho}_r$, exclusively drive the enhancement or suppression of the scalar power spectrum respectively when $c$ is positive or negative. This is a general result that we observed repeatedly in this section: $G(Q)$ only depends on the dissipation strength $Q$ and the power-law temperature dependence in the dissipation rate $c$ which, in the strong dissipative regime, uniquely drives the enhancement or the suppression of the scalar power spectrum.

\subsection{Robustness of the analytical estimate of the scalar power spectrum \label{section:results_analytic}}

\begin{figure}[hb!]
    \centering
    \includegraphics[width=1\linewidth]{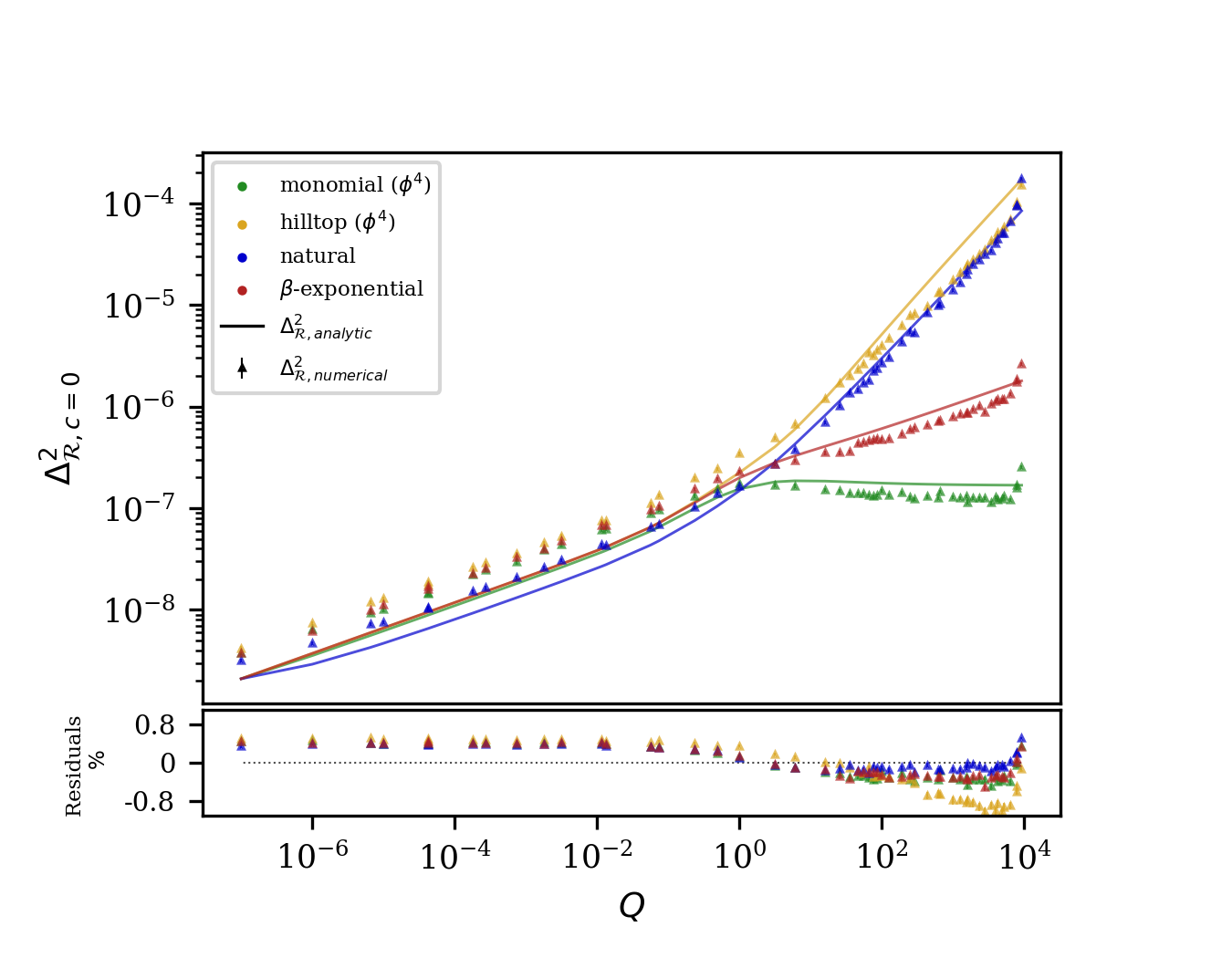}
    \caption{{\it Upper panel:} The scalar power spectrum for $c=0$ as a function of the dissipation strength for different models for the inflaton potential: quartic (green), hilltop (yellow), natural inflation (blue), and $\beta$-exponential (red), see text for the choice of the parameters. Results are shown for the analytical expression in eq.~\eqref{eq:DeltaR_analytic} (solid lines) and for the computation with the numerical code (triangle marks). {\it Lower panel:} The fractional residuals from the two methods over the range of the dissipation strength $Q$ considered.In summary, this figure demonstrates that the analytical estimate to the scalar power spectrum reproduces quite well the numerical results over the entire range of $Q$ considered, with residuals always below $\mathcal{O}(1)$.}
    \label{fig:7}
\end{figure}

We quantify the precision of the analytical estimate of the scalar power spectrum from eq.~\eqref{eq:DeltaR_analytic} against our numerical solutions for the case in which the dissipation rate $\Upsilon$ does not depend on temperature and the inflaton field, i.e.\ $\{c,m\}=\{0,0\}$. Figure~\ref{fig:7} shows the analytical estimate and the numerical solution of the scalar power spectrum for different choices of the inflaton potential, normalized at $Q=10^{-7}$ to the amplitude measured at recombination, namely $\Delta^2_{\mathcal{R}}(k_*)\simeq 2.1\times10^{-9}$ with $k_*=0.05{\rm\,Mpc}^{-1}$~\cite{Planck:2018vyg}.

Overall, the analytical approximation works well for the entire range of dissipation strength $Q$ considered. Numerically, this is presented in the lower panel of figure~\ref{fig:7}, which shows the fractional residuals between the two methods for each inflaton potential used, and they are all $\lesssim \mathcal{O}(1)$. Additionally, one can note that the residuals are always positive for $Q\lesssim \mathcal{O}(1)$ and negative otherwise. This indicates that the analytical approximation marginally under(over)estimates the numerical solution of the scalar power spectrum for $Q\lesssim\mathcal{O}(1)$ ($Q\gtrsim\mathcal{O}(1)$). This behaviour is a direct consequence of the two main assumptions that allowed us to obtain the analytical form of the scalar power spectrum in eq.~\eqref{eq:DeltaR_analytic}, namely: (1) we assumed the temperature $T$ and Hubble parameter $H$ to be constant with respect to $z$ in order to bring them outside of the integral in eq.~\eqref{eq:dphi2}; (2) we neglected the term proportional to the slow-roll coefficient $\eta$ by taking $\alpha\approx\nu$ as defined in eqs.~\eqref{eq:nu} and~\eqref{eq:alpha}.\footnote{Note, here we are comparing the analytical approximation from eq.~\eqref{eq:DeltaR_analytic} to the numerical solution for the case of both $c$ and $m$ equal to zero. Therefore, we are implicitly taking the slow-roll parameter $\beta=0$ in the following discussion.} With respect to assumption (1), $T$ and $H$ are actually both monotonically increasing functions of $z$ and therefore by computing the scalar power spectrum with respect to their value at $z=0$, we are generally always underestimating the true value of the scalar power spectrum. This is particularly relevant for the quantum noise term whose amplitude is quadratically proportional to $H$. Thus, in the weak dissipative regime, i.e.\ for $Q\lesssim 1$, when the quantum noise term still dictates the overall amplitude of the scalar power spectrum, the analytical approximation underestimates the value of the numerical solution, in agreement with our findings. On the other hand, assumption (2) is robust in the weak dissipative regime but becomes obsolete in the strong dissipative regime, for $Q\gg1$, when the slow-roll parameter $\eta$ can be of $\mathcal{O}(Q)$. In this case, one can easily show that the approximated solution in eq.~\eqref{eq:gamma_expression} to the integral in eq.~\eqref{eq:dphi2}, obtained by setting $\alpha\approx\nu$, overestimates the value of the integral when $\alpha\neq\nu $. Particularly, in the strong dissipative regime, this effect is greater than the underestimation  of the temperature $T$ in the thermal noise term arising from assumption (1). Thus, overall from the above argument, we expect the analytical approximation to slightly overestimate the value of the numerical solution for $Q\gg1$, which is again in agreement with our findings.

\section{Summary \& Conclusion}

In this paper we presented \texttt{WarmSPy} which consists—to our knowledge—the first open-source and publicly available code to compute the full, numerical scalar power spectrum for a given warm inflationary (WI) scenario and the scalar dissipation function $G(Q)$ defined in eq.~\eqref{eq:GQ_definition}.
The code is available at \href{https://github.com/GabrieleMonte/WarmSPy.git}{github.com/GabrieleMonte/WarmSPy.git}. The stochastic nature of the problem is faced via a Monte Carlo approach that accounts for the thermal and quantum noise terms. We use the formalism above to assess the dependence of $G(Q)$ on various model parameters and inflationary histories, as well as the effects of metric perturbations. We find that $G(Q)$ only depends on (1) the dissipation strength $Q$, and (2) the dissipation rate's power-law temperature dependence $c$, see eq.~\eqref{eq:Upsilon}, which in the strong dissipative regime $Q\gg 1$ uniquely drives the enhancement or suppression of the scalar power spectrum. In this sense, the scalar dissipation function $G(Q)$ can be thought of as an universal function: for a given value of $c$, $G(Q)$ effectively describes the modification to the power spectrum in eq.~\eqref{eq:defG} regardless of the inflaton model implemented, its history, and possible additional inflaton field dependencies in the dissipation rate or addition of metric perturbations.

Throughout this work, we specifically set $c\in\{3,1,-1\}$ motivated by the explicit constructions of dissipative rates in the literature. For all these cases, we report in table~\ref{tab:1} the best-fit parameters for the proposed fitting functions of $G(Q)$, respectively eqs.~\eqref{eq:GQ_log} and~\eqref{eq:GQ_pol} for $c\in\{3,1\}$ and $c=-1$. Compared to previous work in the literature, the fits we present are valid in a much broader range of the dissipation strength, namely for $Q\in[10^{-7},10^{4}]$, with a minimal number of free parameters. The robustness and broader validity of our numerical fits for the function $G(Q)$ will contribute to phenomenological studies of WI, and specifically help in precisely testing WI models against current and future CMB bounds.

\section*{Acknowledgements}
The authors thank Rudnei O.\ Ramos, Guillermo Ballesteros, Alejandro P.\ Rodriguez, Mathias Pierre and 
Umang Kumar for helpful discussions. The authors would like to make a particular mention to Rudnei O.\ Ramos for their  insightful comments and meticulous review of the manuscript which greatly improved the clarity and quality of our work. K.F.\ is Jeff \& Gail Kodosky Endowed Chair in Physics at the University of Texas at Austin, and K.F.\ and G.M.\ are grateful for support via this Chair. K.F.\ and G.M.\ acknowledge support by the U.S.\ Department of Energy, Office of Science, Office of High Energy Physics program under Award Number DE-SC-0022021 as well as support from the Swedish Research Council (Contract No.~638-2013-8993). V.A.\ acknowledges support from the National Science Foundation under grant number PHY-1914679. L.V.\ and G.M.\ acknowledge respectively the Galileo Galilei Institute for Theoretical Physics in Florence, the INFN Laboratori Nazionali di Frascati and the Oskar Klein Centre in Stockholm University, for their hospitality during the completion of this work. Finally, we would like to acknowledge the Texas Advanced Computing Center (TACC) for providing computational resources and support. The program was tested and all calculations were computed on the TACC infrastructure.

\bibliographystyle{JHEP}
\bibliography{bibl.bib}

\providecommand{\href}[2]{#2}\begingroup\raggedright\begin{thebibliography}{10}

\bibitem{Guth:1980zm}
A.H.~Guth, \emph{{The Inflationary Universe: A Possible Solution to the Horizon
  and Flatness Problems}},
  \href{https://doi.org/10.1103/PhysRevD.23.347}{\emph{Phys. Rev. D} {\bfseries
  23} (1981) 347}.

\bibitem{Linde:1981mu}
A.D.~Linde, \emph{{A New Inflationary Universe Scenario: A Possible Solution of
  the Horizon, Flatness, Homogeneity, Isotropy and Primordial Monopole
  Problems}}, \href{https://doi.org/10.1016/0370-2693(82)91219-9}{\emph{Phys.
  Lett. B} {\bfseries 108} (1982) 389}.

\bibitem{Albrecht:1982wi}
A.~Albrecht and P.J.~Steinhardt, \emph{{Cosmology for Grand Unified Theories
  with Radiatively Induced Symmetry Breaking}},
  \href{https://doi.org/10.1103/PhysRevLett.48.1220}{\emph{Phys. Rev. Lett.}
  {\bfseries 48} (1982) 1220}.

\bibitem{Kazanas:1980tx}
D.~Kazanas, \emph{{Dynamics of the Universe and Spontaneous Symmetry
  Breaking}}, \href{https://doi.org/10.1086/183361}{\emph{Astrophys. J. Lett.}
  {\bfseries 241} (1980) L59}.

\bibitem{Starobinsky:1980te}
A.A.~Starobinsky, \emph{{A New Type of Isotropic Cosmological Models Without
  Singularity}},
  \href{https://doi.org/10.1016/0370-2693(80)90670-X}{\emph{Phys. Lett. B}
  {\bfseries 91} (1980) 99}.

\bibitem{Sato:1980yn}
K.~Sato, \emph{{First Order Phase Transition of a Vacuum and Expansion of the
  Universe}}, {\emph{Mon. Not. Roy. Astron. Soc.} {\bfseries 195} (1981) 467}.

\bibitem{Mukhanov:1981xt}
V.F.~Mukhanov and G.V.~Chibisov, \emph{{Quantum Fluctuations and a Nonsingular
  Universe}}, {\emph{JETP Lett.} {\bfseries 33} (1981) 532}.

\bibitem{Linde:1983gd}
A.D.~Linde, \emph{{Chaotic Inflation}},
  \href{https://doi.org/10.1016/0370-2693(83)90837-7}{\emph{Phys. Lett. B}
  {\bfseries 129} (1983) 177}.

\bibitem{Mukhanov:1990me}
V.F.~Mukhanov, H.A.~Feldman and R.H.~Brandenberger, \emph{{Theory of
  cosmological perturbations. Part 1. Classical perturbations. Part 2. Quantum
  theory of perturbations. Part 3. Extensions}},
  \href{https://doi.org/10.1016/0370-1573(92)90044-Z}{\emph{Phys. Rept.}
  {\bfseries 215} (1992) 203}.

\bibitem{Planck:2018nkj}
{\scshape Planck} collaboration, \emph{{Planck 2018 results. I. Overview and
  the cosmological legacy of Planck}},
  \href{https://doi.org/10.1051/0004-6361/201833880}{\emph{Astron. Astrophys.}
  {\bfseries 641} (2020) A1}
  [\href{https://arxiv.org/abs/1807.06205}{{\ttfamily 1807.06205}}].

\bibitem{Planck:2018jri}
{\scshape Planck} collaboration, \emph{{Planck 2018 results. X. Constraints on
  inflation}}, \href{https://doi.org/10.1051/0004-6361/201833887}{\emph{Astron.
  Astrophys.} {\bfseries 641} (2020) A10}
  [\href{https://arxiv.org/abs/1807.06211}{{\ttfamily 1807.06211}}].

\bibitem{ACT:2020gnv}
{\scshape ACT} collaboration, \emph{{The Atacama Cosmology Telescope: DR4 Maps
  and Cosmological Parameters}},
  \href{https://doi.org/10.1088/1475-7516/2020/12/047}{\emph{JCAP} {\bfseries
  12} (2020) 047} [\href{https://arxiv.org/abs/2007.07288}{{\ttfamily
  2007.07288}}].

\bibitem{ACT:2020frw}
{\scshape ACT} collaboration, \emph{{The Atacama Cosmology Telescope: a
  measurement of the Cosmic Microwave Background power spectra at 98 and 150
  GHz}}, \href{https://doi.org/10.1088/1475-7516/2020/12/045}{\emph{JCAP}
  {\bfseries 12} (2020) 045}
  [\href{https://arxiv.org/abs/2007.07289}{{\ttfamily 2007.07289}}].

\bibitem{ACT:2023dou}
{\scshape ACT} collaboration, \emph{{The Atacama Cosmology Telescope: A
  Measurement of the DR6 CMB Lensing Power Spectrum and its Implications for
  Structure Growth}},  \href{https://arxiv.org/abs/2304.05202}{{\ttfamily
  2304.05202}}.

\bibitem{SPT-3G:2021eoc}
{\scshape SPT-3G} collaboration, \emph{{Measurements of the E-mode polarization
  and temperature-E-mode correlation of the CMB from SPT-3G 2018 data}},
  \href{https://doi.org/10.1103/PhysRevD.104.022003}{\emph{Phys. Rev. D}
  {\bfseries 104} (2021) 022003}
  [\href{https://arxiv.org/abs/2101.01684}{{\ttfamily 2101.01684}}].

\bibitem{SPT-3G:2021wgf}
{\scshape SPT-3G} collaboration, \emph{{Constraints on \ensuremath{\Lambda}CDM
  extensions from the SPT-3G 2018 EE and TE power spectra}},
  \href{https://doi.org/10.1103/PhysRevD.104.083509}{\emph{Phys. Rev. D}
  {\bfseries 104} (2021) 083509}
  [\href{https://arxiv.org/abs/2103.13618}{{\ttfamily 2103.13618}}].

\bibitem{Berera:1995ie}
A.~Berera, \emph{{Warm inflation}},
  \href{https://doi.org/10.1103/PhysRevLett.75.3218}{\emph{Phys. Rev. Lett.}
  {\bfseries 75} (1995) 3218}
  [\href{https://arxiv.org/abs/astro-ph/9509049}{{\ttfamily
  astro-ph/9509049}}].

\bibitem{Berera:1996fm}
A.~Berera, \emph{{Interpolating the stage of exponential expansion in the early
  universe: A Possible alternative with no reheating}},
  \href{https://doi.org/10.1103/PhysRevD.55.3346}{\emph{Phys. Rev. D}
  {\bfseries 55} (1997) 3346}
  [\href{https://arxiv.org/abs/hep-ph/9612239}{{\ttfamily hep-ph/9612239}}].

\bibitem{Fang:1980wi}
L.Z.~Fang, \emph{{Entropy Generation in the Early Universe by Dissipative
  Processes Near the Higgs' Phase Transitions}},
  \href{https://doi.org/10.1016/0370-2693(80)90421-9}{\emph{Phys. Lett. B}
  {\bfseries 95} (1980) 154}.

\bibitem{Moss:1985wn}
I.G.~Moss, \emph{{Primordial Inflation With Spontaneous Symmetry Breaking}},
  \href{https://doi.org/10.1016/0370-2693(85)90570-2}{\emph{Phys. Lett. B}
  {\bfseries 154} (1985) 120}.

\bibitem{Yokoyama:1987an}
J.~Yokoyama and K.-i.~Maeda, \emph{{On the Dynamics of the Power Law Inflation
  Due to an Exponential Potential}},
  \href{https://doi.org/10.1016/0370-2693(88)90880-5}{\emph{Phys. Lett. B}
  {\bfseries 207} (1988) 31}.

\bibitem{Berera:2008ar}
A.~Berera, I.G.~Moss and R.O.~Ramos, \emph{{Warm Inflation and its
  Microphysical Basis}},
  \href{https://doi.org/10.1088/0034-4885/72/2/026901}{\emph{Rept. Prog. Phys.}
  {\bfseries 72} (2009) 026901}
  [\href{https://arxiv.org/abs/0808.1855}{{\ttfamily 0808.1855}}].

\bibitem{Bastero-Gil:2009sdq}
M.~Bastero-Gil and A.~Berera, \emph{{Warm inflation model building}},
  \href{https://doi.org/10.1142/S0217751X09044206}{\emph{Int. J. Mod. Phys. A}
  {\bfseries 24} (2009) 2207}
  [\href{https://arxiv.org/abs/0902.0521}{{\ttfamily 0902.0521}}].

\bibitem{Kamali:2023lzq}
V.~Kamali, M.~Motaharfar and R.~O.~Ramos, \emph{{Recent Developments in Warm
  Inflation}}, \href{https://doi.org/10.3390/universe9030124}{\emph{Universe}
  {\bfseries 9} (2023) 124} [\href{https://arxiv.org/abs/2302.02827}{{\ttfamily
  2302.02827}}].

\bibitem{Ramos:2013nsa}
R.O.~Ramos and L.A.~da~Silva, \emph{{Power spectrum for inflation models with
  quantum and thermal noises}},
  \href{https://doi.org/10.1088/1475-7516/2013/03/032}{\emph{JCAP} {\bfseries
  03} (2013) 032} [\href{https://arxiv.org/abs/1302.3544}{{\ttfamily
  1302.3544}}].

\bibitem{Hall:2003zp}
L.M.H.~Hall, I.G.~Moss and A.~Berera, \emph{{Scalar perturbation spectra from
  warm inflation}},
  \href{https://doi.org/10.1103/PhysRevD.69.083525}{\emph{Phys. Rev. D}
  {\bfseries 69} (2004) 083525}
  [\href{https://arxiv.org/abs/astro-ph/0305015}{{\ttfamily
  astro-ph/0305015}}].

\bibitem{Graham:2009bf}
C.~Graham and I.G.~Moss, \emph{{Density fluctuations from warm inflation}},
  \href{https://doi.org/10.1088/1475-7516/2009/07/013}{\emph{JCAP} {\bfseries
  07} (2009) 013} [\href{https://arxiv.org/abs/0905.3500}{{\ttfamily
  0905.3500}}].

\bibitem{Bastero-Gil:2011rva}
M.~Bastero-Gil, A.~Berera and R.O.~Ramos, \emph{{Shear viscous effects on the
  primordial power spectrum from warm inflation}},
  \href{https://doi.org/10.1088/1475-7516/2011/07/030}{\emph{JCAP} {\bfseries
  07} (2011) 030} [\href{https://arxiv.org/abs/1106.0701}{{\ttfamily
  1106.0701}}].

\bibitem{huston:2011vt}
I.~Huston and K.A.~Malik, \emph{{Second Order Perturbations During Inflation
  Beyond Slow-roll}},
  \href{https://doi.org/10.1088/1475-7516/2011/10/029}{\emph{JCAP} {\bfseries
  10} (2011) 029} [\href{https://arxiv.org/abs/1103.0912}{{\ttfamily
  1103.0912}}].

\bibitem{rosati_robert_2020_4708348}
R.~Rosati, \emph{{Inflation.jl -- A Julia package for numerical evaluation of
  cosmic inflation models using the transport method}},  July, 2020.
\newblock 10.5281/zenodo.4708348.

\bibitem{price:2014xpa}
L.C.~Price, J.~Frazer, J.~Xu, H.V.~Peiris and R.~Easther, \emph{{MultiModeCode:
  An efficient numerical solver for multifield inflation}},
  \href{https://doi.org/10.1088/1475-7516/2015/03/005}{\emph{JCAP} {\bfseries
  03} (2015) 005} [\href{https://arxiv.org/abs/1410.0685}{{\ttfamily
  1410.0685}}].

\bibitem{seery:2016lko}
D.~Seery, \emph{{CppTransport: a platform to automate calculation of
  inflationary correlation functions}},
  \href{https://arxiv.org/abs/1609.00380}{{\ttfamily 1609.00380}}.

\bibitem{mulryne:2016mzv}
D.J.~Mulryne and J.W.~Ronayne, \emph{{PyTransport: A Python package for the
  calculation of inflationary correlation functions}},
  \href{https://doi.org/10.21105/joss.00494}{\emph{J. Open Source Softw.}
  {\bfseries 3} (2018) 494} [\href{https://arxiv.org/abs/1609.00381}{{\ttfamily
  1609.00381}}].

\bibitem{dias:2016rjq}
M.~Dias, J.~Frazer, D.J.~Mulryne and D.~Seery, \emph{{Numerical evaluation of
  the bispectrum in multiple field inflation\textemdash{}the transport approach
  with code}}, \href{https://doi.org/10.1088/1475-7516/2016/12/033}{\emph{JCAP}
  {\bfseries 12} (2016) 033}
  [\href{https://arxiv.org/abs/1609.00379}{{\ttfamily 1609.00379}}].

\bibitem{ronayne:2017qzn}
J.W.~Ronayne and D.J.~Mulryne, \emph{{Numerically evaluating the bispectrum in
  curved field-space\textemdash{} with PyTransport 2.0}},
  \href{https://doi.org/10.1088/1475-7516/2018/01/023}{\emph{JCAP} {\bfseries
  01} (2018) 023} [\href{https://arxiv.org/abs/1708.07130}{{\ttfamily
  1708.07130}}].

\bibitem{hazra:2012yn}
D.K.~Hazra, L.~Sriramkumar and J.~Martin, \emph{{BINGO: A code for the
  efficient computation of the scalar bi-spectrum}},
  \href{https://doi.org/10.1088/1475-7516/2013/05/026}{\emph{JCAP} {\bfseries
  05} (2013) 026} [\href{https://arxiv.org/abs/1201.0926}{{\ttfamily
  1201.0926}}].

\bibitem{chen:2006xjb}
X.~Chen, R.~Easther and E.A.~Lim, \emph{{Large Non-Gaussianities in Single
  Field Inflation}},
  \href{https://doi.org/10.1088/1475-7516/2007/06/023}{\emph{JCAP} {\bfseries
  06} (2007) 023} [\href{https://arxiv.org/abs/astro-ph/0611645}{{\ttfamily
  astro-ph/0611645}}].

\bibitem{horner:2013sea}
J.S.~Horner and C.R.~Contaldi, \emph{{Non-Gaussian signatures of general
  inflationary trajectories}},
  \href{https://doi.org/10.1088/1475-7516/2014/09/001}{\emph{JCAP} {\bfseries
  09} (2014) 001} [\href{https://arxiv.org/abs/1311.3224}{{\ttfamily
  1311.3224}}].

\bibitem{Das:2020xmh}
S.~Das and R.O.~Ramos, \emph{{Runaway potentials in warm inflation satisfying
  the swampland conjectures}},
  \href{https://doi.org/10.1103/PhysRevD.102.103522}{\emph{Phys. Rev. D}
  {\bfseries 102} (2020) 103522}
  [\href{https://arxiv.org/abs/2007.15268}{{\ttfamily 2007.15268}}].

\bibitem{Berghaus:2019whh}
K.V.~Berghaus, P.W.~Graham and D.E.~Kaplan, \emph{{Minimal Warm Inflation}},
  \href{https://doi.org/10.1088/1475-7516/2020/03/034}{\emph{JCAP} {\bfseries
  03} (2020) 034} [\href{https://arxiv.org/abs/1910.07525}{{\ttfamily
  1910.07525}}].

\bibitem{Bastero-Gil:2019gao}
M.~Bastero-Gil, A.~Berera, R.O.~Ramos and J.G.~Rosa, \emph{{Towards a reliable
  effective field theory of inflation}},
  \href{https://doi.org/10.1016/j.physletb.2020.136055}{\emph{Phys. Lett. B}
  {\bfseries 813} (2021) 136055}
  [\href{https://arxiv.org/abs/1907.13410}{{\ttfamily 1907.13410}}].

\bibitem{Berera:1995wh}
A.~Berera and L.-Z.~Fang, \emph{{Thermally induced density perturbations in the
  inflation era}},
  \href{https://doi.org/10.1103/PhysRevLett.74.1912}{\emph{Phys. Rev. Lett.}
  {\bfseries 74} (1995) 1912}
  [\href{https://arxiv.org/abs/astro-ph/9501024}{{\ttfamily
  astro-ph/9501024}}].

\bibitem{Berera:1998gx}
A.~Berera, M.~Gleiser and R.O.~Ramos, \emph{{Strong dissipative behavior in
  quantum field theory}},
  \href{https://doi.org/10.1103/PhysRevD.58.123508}{\emph{Phys. Rev. D}
  {\bfseries 58} (1998) 123508}
  [\href{https://arxiv.org/abs/hep-ph/9803394}{{\ttfamily hep-ph/9803394}}].

\bibitem{Yokoyama:1998ju}
J.~Yokoyama and A.D.~Linde, \emph{{Is warm inflation possible?}},
  \href{https://doi.org/10.1103/PhysRevD.60.083509}{\emph{Phys. Rev. D}
  {\bfseries 60} (1999) 083509}
  [\href{https://arxiv.org/abs/hep-ph/9809409}{{\ttfamily hep-ph/9809409}}].

\bibitem{Berera:2002sp}
A.~Berera and R.O.~Ramos, \emph{{Construction of a robust warm inflation
  mechanism}},
  \href{https://doi.org/10.1016/j.physletb.2003.06.028}{\emph{Phys. Lett. B}
  {\bfseries 567} (2003) 294}
  [\href{https://arxiv.org/abs/hep-ph/0210301}{{\ttfamily hep-ph/0210301}}].

\bibitem{Moss:2006gt}
I.G.~Moss and C.~Xiong, \emph{{Dissipation coefficients for supersymmetric
  inflatonary models}},  \href{https://arxiv.org/abs/hep-ph/0603266}{{\ttfamily
  hep-ph/0603266}}.

\bibitem{Bastero-Gil:2016qru}
M.~Bastero-Gil, A.~Berera, R.O.~Ramos and J.G.~Rosa, \emph{{Warm Little
  Inflaton}}, \href{https://doi.org/10.1103/PhysRevLett.117.151301}{\emph{Phys.
  Rev. Lett.} {\bfseries 117} (2016) 151301}
  [\href{https://arxiv.org/abs/1604.08838}{{\ttfamily 1604.08838}}].

\bibitem{Bastero-Gil:2010dgy}
M.~Bastero-Gil, A.~Berera and R.O.~Ramos, \emph{{Dissipation coefficients from
  scalar and fermion quantum field interactions}},
  \href{https://doi.org/10.1088/1475-7516/2011/09/033}{\emph{JCAP} {\bfseries
  09} (2011) 033} [\href{https://arxiv.org/abs/1008.1929}{{\ttfamily
  1008.1929}}].

\bibitem{Moore:2010jd}
G.D.~Moore and M.~Tassler, \emph{{The Sphaleron Rate in SU(N) Gauge Theory}},
  \href{https://doi.org/10.1007/JHEP02(2011)105}{\emph{JHEP} {\bfseries 02}
  (2011) 105} [\href{https://arxiv.org/abs/1011.1167}{{\ttfamily 1011.1167}}].

\bibitem{Laine:2016hma}
M.~Laine and A.~Vuorinen, \emph{{Basics of Thermal Field Theory}}, vol.~925,
  Springer (2016),
  \href{https://doi.org/10.1007/978-3-319-31933-9}{10.1007/978-3-319-31933-9},
  [\href{https://arxiv.org/abs/1701.01554}{{\ttfamily 1701.01554}}].

\bibitem{Berghaus:2020ekh}
K.V.~Berghaus, P.W.~Graham, D.E.~Kaplan, G.D.~Moore and S.~Rajendran,
  \emph{{Dark energy radiation}},
  \href{https://doi.org/10.1103/PhysRevD.104.083520}{\emph{Phys. Rev. D}
  {\bfseries 104} (2021) 083520}
  [\href{https://arxiv.org/abs/2012.10549}{{\ttfamily 2012.10549}}].

\bibitem{Montefalcone:2022jfw}
G.~Montefalcone, V.~Aragam, L.~Visinelli and K.~Freese, \emph{{Observational
  constraints on warm natural inflation}},
  \href{https://doi.org/10.1088/1475-7516/2023/03/002}{\emph{JCAP} {\bfseries
  03} (2023) 002} [\href{https://arxiv.org/abs/2212.04482}{{\ttfamily
  2212.04482}}].

\bibitem{Taylor:2000ze}
A.N.~Taylor and A.~Berera, \emph{{Perturbation spectra in the warm inflationary
  scenario}}, \href{https://doi.org/10.1103/PhysRevD.62.083517}{\emph{Phys.
  Rev. D} {\bfseries 62} (2000) 083517}
  [\href{https://arxiv.org/abs/astro-ph/0006077}{{\ttfamily
  astro-ph/0006077}}].

\bibitem{Bastero-Gil:2004oun}
M.~Bastero-Gil and A.~Berera, \emph{{Determining the regimes of cold and warm
  inflation in the SUSY hybrid model}},
  \href{https://doi.org/10.1103/PhysRevD.71.063515}{\emph{Phys. Rev. D}
  {\bfseries 71} (2005) 063515}
  [\href{https://arxiv.org/abs/hep-ph/0411144}{{\ttfamily hep-ph/0411144}}].

\bibitem{Moss:2008yb}
I.G.~Moss and C.~Xiong, \emph{{On the consistency of warm inflation}},
  \href{https://doi.org/10.1088/1475-7516/2008/11/023}{\emph{JCAP} {\bfseries
  11} (2008) 023} [\href{https://arxiv.org/abs/0808.0261}{{\ttfamily
  0808.0261}}].

\bibitem{Ballesteros:2023dno}
G.~Ballesteros, A.~P\'erez~Rodr\'\i{}guez and M.~Pierre, \emph{{Monomial warm
  inflation revisited}},  \href{https://arxiv.org/abs/2304.05978}{{\ttfamily
  2304.05978}}.

\bibitem{Ballesteros:2022hjk}
G.~Ballesteros, M.A.G.~Garc\'\i{}a, A.P.~Rodr\'\i{}guez, M.~Pierre and J.~Rey,
  \emph{{Primordial black holes and gravitational waves from dissipation during
  inflation}}, \href{https://doi.org/10.1088/1475-7516/2022/12/006}{\emph{JCAP}
  {\bfseries 12} (2022) 006}
  [\href{https://arxiv.org/abs/2208.14978}{{\ttfamily 2208.14978}}].

\bibitem{Starobinsky:1986fx}
A.A.~Starobinsky, \emph{{STOCHASTIC DE SITTER (INFLATIONARY) STAGE IN THE EARLY
  UNIVERSE}}, \href{https://doi.org/10.1007/3-540-16452-9_6}{\emph{Lect. Notes
  Phys.} {\bfseries 246} (1986) 107}.

\bibitem{Bastero-Gil:2014jsa}
M.~Bastero-Gil, A.~Berera, I.G.~Moss and R.O.~Ramos, \emph{{Cosmological
  fluctuations of a random field and radiation fluid}},
  \href{https://doi.org/10.1088/1475-7516/2014/05/004}{\emph{JCAP} {\bfseries
  05} (2014) 004} [\href{https://arxiv.org/abs/1401.1149}{{\ttfamily
  1401.1149}}].

\bibitem{liddle_lyth_2000}
A.R.~Liddle and D.H.~Lyth, \emph{Cosmological Inflation and Large-Scale
  Structure}, Cambridge University Press (2000),
  \href{https://doi.org/10.1017/CBO9781139175180}{10.1017/CBO9781139175180}.

\bibitem{BICEP:2021xfz}
{\scshape BICEP, Keck} collaboration, \emph{{Improved Constraints on Primordial
  Gravitational Waves using Planck, WMAP, and BICEP/Keck Observations through
  the 2018 Observing Season}},
  \href{https://doi.org/10.1103/PhysRevLett.127.151301}{\emph{Phys. Rev. Lett.}
  {\bfseries 127} (2021) 151301}
  [\href{https://arxiv.org/abs/2110.00483}{{\ttfamily 2110.00483}}].

\bibitem{Bhattacharya:2006dm}
K.~Bhattacharya, S.~Mohanty and A.~Nautiyal, \emph{{Enhanced polarization of
  CMB from thermal gravitational waves}},
  \href{https://doi.org/10.1103/PhysRevLett.97.251301}{\emph{Phys. Rev. Lett.}
  {\bfseries 97} (2006) 251301}
  [\href{https://arxiv.org/abs/astro-ph/0607049}{{\ttfamily
  astro-ph/0607049}}].

\bibitem{Qiu:2021ytc}
Y.~Qiu and L.~Sorbo, \emph{{Spectrum of tensor perturbations in warm
  inflation}}, \href{https://doi.org/10.1103/PhysRevD.104.083542}{\emph{Phys.
  Rev. D} {\bfseries 104} (2021) 083542}
  [\href{https://arxiv.org/abs/2107.09754}{{\ttfamily 2107.09754}}].

\bibitem{Namba:2015gja}
R.~Namba, M.~Peloso, M.~Shiraishi, L.~Sorbo and C.~Unal, \emph{{Scale-dependent
  gravitational waves from a rolling axion}},
  \href{https://doi.org/10.1088/1475-7516/2016/01/041}{\emph{JCAP} {\bfseries
  01} (2016) 041} [\href{https://arxiv.org/abs/1509.07521}{{\ttfamily
  1509.07521}}].

\bibitem{Peloso:2016gqs}
M.~Peloso, L.~Sorbo and C.~Unal, \emph{{Rolling axions during inflation:
  perturbativity and signatures}},
  \href{https://doi.org/10.1088/1475-7516/2016/09/001}{\emph{JCAP} {\bfseries
  09} (2016) 001} [\href{https://arxiv.org/abs/1606.00459}{{\ttfamily
  1606.00459}}].

\bibitem{Benetti:2016jhf}
M.~Benetti and R.O.~Ramos, \emph{{Warm inflation dissipative effects:
  predictions and constraints from the Planck data}},
  \href{https://doi.org/10.1103/PhysRevD.95.023517}{\emph{Phys. Rev. D}
  {\bfseries 95} (2017) 023517}
  [\href{https://arxiv.org/abs/1610.08758}{{\ttfamily 1610.08758}}].

\bibitem{Boubekeur:2005zm}
L.~Boubekeur and D.H.~Lyth, \emph{{Hilltop inflation}},
  \href{https://doi.org/10.1088/1475-7516/2005/07/010}{\emph{JCAP} {\bfseries
  07} (2005) 010} [\href{https://arxiv.org/abs/hep-ph/0502047}{{\ttfamily
  hep-ph/0502047}}].

\bibitem{Bartrum:2013oka}
S.~Bartrum, A.~Berera and J.a.G.~Rosa, \emph{{Warming up for Planck}},
  \href{https://doi.org/10.1088/1475-7516/2013/06/025}{\emph{JCAP} {\bfseries
  06} (2013) 025} [\href{https://arxiv.org/abs/1303.3508}{{\ttfamily
  1303.3508}}].

\bibitem{Freese:1990rb}
K.~Freese, J.A.~Frieman and A.V.~Olinto, \emph{{Natural inflation with pseudo -
  Nambu-Goldstone bosons}},
  \href{https://doi.org/10.1103/PhysRevLett.65.3233}{\emph{Phys. Rev. Lett.}
  {\bfseries 65} (1990) 3233}.

\bibitem{Adams:1990pn}
F.C.~Adams, K.~Freese and A.H.~Guth, \emph{{Constraints on the scalar field
  potential in inflationary models}},
  \href{https://doi.org/10.1103/PhysRevD.43.965}{\emph{Phys. Rev. D} {\bfseries
  43} (1991) 965}.

\bibitem{Montefalcone:2022owy}
G.~Montefalcone, V.~Aragam, L.~Visinelli and K.~Freese, \emph{{Constraints on
  the scalar-field potential in warm inflation}},
  \href{https://doi.org/10.1103/PhysRevD.107.063543}{\emph{Phys. Rev. D}
  {\bfseries 107} (2023) 063543}
  [\href{https://arxiv.org/abs/2209.14908}{{\ttfamily 2209.14908}}].

\bibitem{Visinelli:2011jy}
L.~Visinelli, \emph{{Natural Warm Inflation}},
  \href{https://doi.org/10.1088/1475-7516/2011/09/013}{\emph{JCAP} {\bfseries
  09} (2011) 013} [\href{https://arxiv.org/abs/1107.3523}{{\ttfamily
  1107.3523}}].

\bibitem{Mishra:2011vh}
H.~Mishra, S.~Mohanty and A.~Nautiyal, \emph{{Warm natural inflation}},
  \href{https://doi.org/10.1016/j.physletb.2012.02.005}{\emph{Phys. Lett. B}
  {\bfseries 710} (2012) 245}
  [\href{https://arxiv.org/abs/1106.3039}{{\ttfamily 1106.3039}}].

\bibitem{Kamali:2019ppi}
V.~Kamali, \emph{{Warm pseudoscalar inflation}},
  \href{https://doi.org/10.1103/PhysRevD.100.043520}{\emph{Phys. Rev. D}
  {\bfseries 100} (2019) 043520}
  [\href{https://arxiv.org/abs/1901.01897}{{\ttfamily 1901.01897}}].

\bibitem{Correa:2022ngq}
M.~Correa, M.R.~Gangopadhyay, N.~Jaman and G.J.~Mathews, \emph{{Primordial
  black-hole dark matter via warm natural inflation}},
  \href{https://doi.org/10.1016/j.physletb.2022.137510}{\emph{Phys. Lett. B}
  {\bfseries 835} (2022) 137510}
  [\href{https://arxiv.org/abs/2207.10394}{{\ttfamily 2207.10394}}].

\bibitem{Alcaniz:2006nu}
J.S.~Alcaniz and F.C.~Carvalho, \emph{{Beta-exponential inflation}},
  \href{https://doi.org/10.1209/0295-5075/79/39001}{\emph{EPL} {\bfseries 79}
  (2007) 39001} [\href{https://arxiv.org/abs/astro-ph/0612279}{{\ttfamily
  astro-ph/0612279}}].

\bibitem{dosSantos:2021vis}
F.B.M.~dos Santos, S.~Santos~da Costa, R.~Silva, M.~Benetti and J.~Alcaniz,
  \emph{{Constraining non-minimally coupled \ensuremath{\beta}-exponential
  inflation with CMB data}},
  \href{https://doi.org/10.1088/1475-7516/2022/06/001}{\emph{JCAP} {\bfseries
  06} (2022) 001} [\href{https://arxiv.org/abs/2110.14758}{{\ttfamily
  2110.14758}}].

\bibitem{Santos:2022exm}
F.B.M.d.~Santos, R.~Silva, S.S.~da~Costa, M.~Benetti and J.S.~Alcaniz,
  \emph{{Warm $\beta $-exponential inflation and the swampland conjectures}},
  \href{https://doi.org/10.1140/epjc/s10052-023-11329-w}{\emph{Eur. Phys. J. C}
  {\bfseries 83} (2023) 178}
  [\href{https://arxiv.org/abs/2209.06153}{{\ttfamily 2209.06153}}].

\bibitem{Weinberg:2008zzc}
S.~Weinberg, \emph{{Cosmology}}, Oxford University Press, Oxford UK (2008).

\bibitem{Motaharfar:2018zyb}
M.~Motaharfar, V.~Kamali and R.O.~Ramos, \emph{{Warm inflation as a way out of
  the swampland}},
  \href{https://doi.org/10.1103/PhysRevD.99.063513}{\emph{Phys. Rev. D}
  {\bfseries 99} (2019) 063513}
  [\href{https://arxiv.org/abs/1810.02816}{{\ttfamily 1810.02816}}].

\bibitem{Berera:1998px}
A.~Berera, M.~Gleiser and R.O.~Ramos, \emph{{A First principles warm inflation
  model that solves the cosmological horizon / flatness problems}},
  \href{https://doi.org/10.1103/PhysRevLett.83.264}{\emph{Phys. Rev. Lett.}
  {\bfseries 83} (1999) 264}
  [\href{https://arxiv.org/abs/hep-ph/9809583}{{\ttfamily hep-ph/9809583}}].

\bibitem{Bastero-Gil:2012akf}
M.~Bastero-Gil, A.~Berera, R.O.~Ramos and J.G.~Rosa, \emph{{General dissipation
  coefficient in low-temperature warm inflation}},
  \href{https://doi.org/10.1088/1475-7516/2013/01/016}{\emph{JCAP} {\bfseries
  01} (2013) 016} [\href{https://arxiv.org/abs/1207.0445}{{\ttfamily
  1207.0445}}].

\bibitem{Planck:2018vyg}
{\scshape Planck} collaboration, \emph{{Planck 2018 results. VI. Cosmological
  parameters}},
  \href{https://doi.org/10.1051/0004-6361/201833910}{\emph{Astron. Astrophys.}
  {\bfseries 641} (2020) A6}
  [\href{https://arxiv.org/abs/1807.06209}{{\ttfamily 1807.06209}}].

\end{thebibliography}\endgroup
\clearpage
\appendix
\section{Additional plots \label{Appendix:plots}}
\begin{figure}[hb!]
    \centering
    \includegraphics[width=\linewidth]{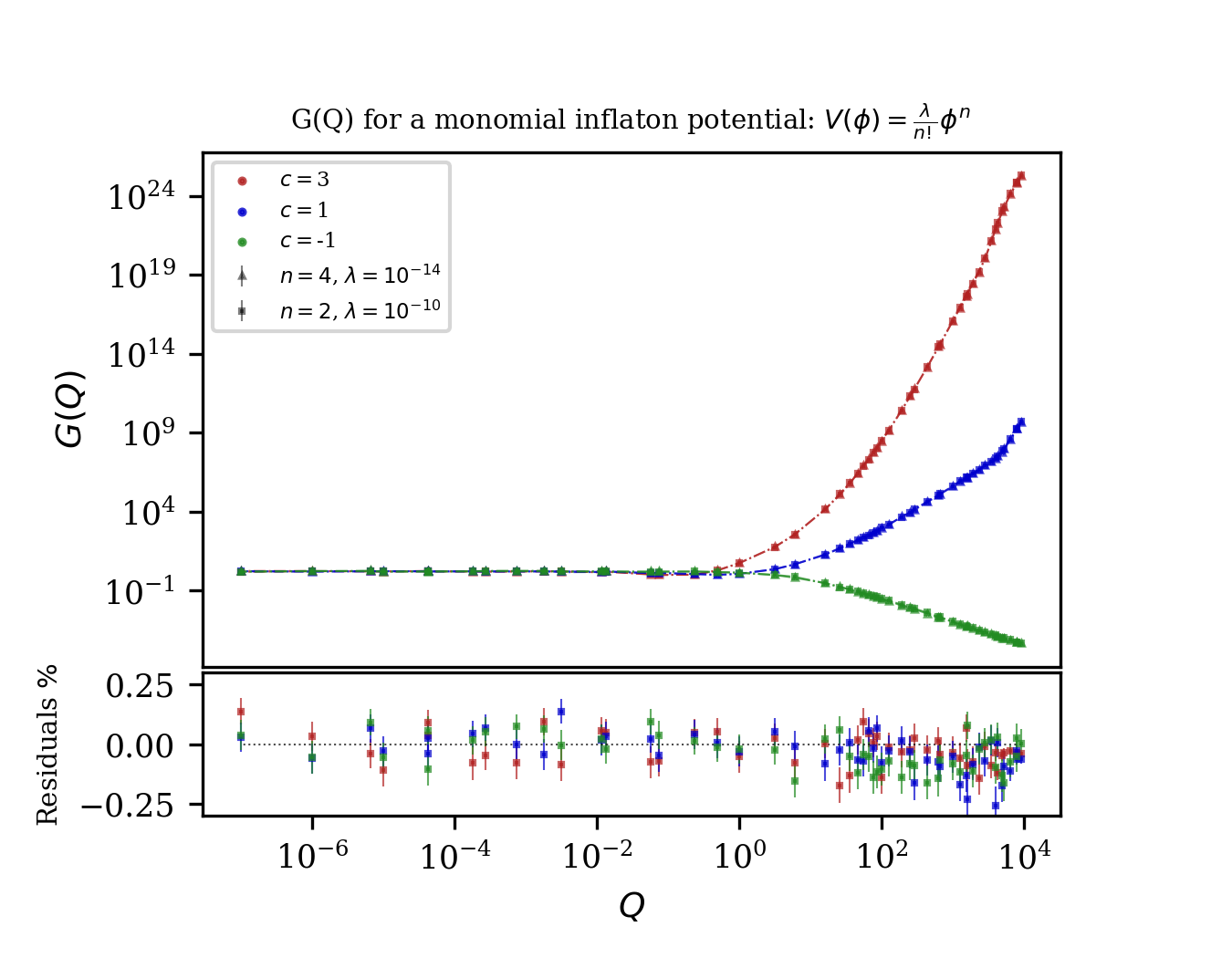}
    \caption{{\it Upper panel:} The scalar dissipation function $G(Q)$ for a monomial potential, see eq.~\eqref{eq:monomial_potential}, with different values for its parameters $n$ and $\lambda$: $\{n,\lambda\}=\{4,10^{-14}\}$ (triangle); $\{n,\lambda\}=\{2,10^{-10}\}$ (squares). Results are shown for a cubic, linear and inverse temperature dependence in the dissipation rate, respectively in red, blue and green. {\it Lower panel:} The fractional residuals from the two different choices of inflaton potential parameters, over the range of the dissipation strength $Q$ considered. In short, this figure demonstrates that for a given form of the inflaton potential, $G(Q)$ is robust to changes in the potential parameters.}
    \label{fig:A1}
\end{figure}
\section{Automated initial conditions for the \texttt{Background} module
\label{Appendix:ICs}}
In order to solve the perturbations' equations of motion, we must first determine the initial condition for the inflaton field $\phi_0$ that guarantees $N^{\rm{end}}_e$ $e$-folds before inflation ends. We first define the function \texttt{Bg\_solver\_test} in the \texttt{Background} module. This is an iterative solver of the background evolution that, starting from an initial guess $\phi_0^{\rm{guess}}$, scans different values of $\phi$ via a standard gradient descent algorithm until the condition $\epsilon_H (N^{\rm{end}}_e ) = 1$ is satisfied. At each iterative step, the initial value of the inflaton field $\phi_0$ is displaced according to:
\begin{align}
    \delta\phi_0&= \frac{d \epsilon_H}{d \phi}\Big|_{\phi=\phi_0} \left[1-\epsilon_H(N_e^{\rm{end}})\right] \cdot l_r, \\
    &\simeq - \frac{V_{,\phi} \epsilon_H}{V}\Big|_{\phi=\phi_0} \left[1-\epsilon_H(N_e^{\rm{end}})\right]\cdot l_r,
\end{align}
where the last line follows from assuming slow-roll, i.e.\ $d\epsilon_H/d\phi\simeq-\epsilon_H V/V_{,\phi}$.
$l_r$ is the gradient descent learning rate, whose value is chosen to ensure convergence of the gradient descent algorithm and determines the overall step size taken in each iteration to update $\phi_0$. Depending on the inflaton model, the optimal value of $l_r$ can vary significantly. Generally we recommend starting with a small value of $l_r$ to avoid overshooting, and then incrementally increasing it if convergence is too slow.

To determine the initial condition on $\phi$ efficiently, it is useful (when possible) to set $\phi_0^{\rm{guess}}$ to the value obtained by analytically solving the background equations in the slow-roll approximation and for a constant dissipation strength $Q$. We implement this approach via the function \texttt{Phi0\_slow\_roll\_guesser} which outputs the slow-roll guess for $\phi_0$ for a given value of $N_e^{\rm{end}}$ and any of the inflaton potentials available in the code (see section~\ref{sec:Inflaton_Model} for details).

\end{document}